\documentclass[twocolumn]{aastex631}
\usepackage{soul}

\shorttitle{Microwave transients from the quiet solar corona}
\shortauthors{Mondal et al.}

\begin{document}

\title{Multifrequency microwave imaging of weak transients from the quiet solar corona}

\correspondingauthor{Surajit Mondal}
\email{surajit.mondal@njit.edu}

\author[0000-0002-2325-5298]{Surajit Mondal}
\affiliation{Center for Solar-Terrestrial Research, New Jersey Institute of Technology,\\
323 M L King Jr Boulevard, Newark, NJ 07102-1982, USA}

\author[0000-0002-2325-5298]{Bin Chen}
\affiliation{Center for Solar-Terrestrial Research, New Jersey Institute of Technology,\\
323 M L King Jr Boulevard, Newark, NJ 07102-1982, USA}

\author[0000-0002-2325-5298]{Sijie Yu}
\affiliation{Center for Solar-Terrestrial Research, New Jersey Institute of Technology,\\
323 M L King Jr Boulevard, Newark, NJ 07102-1982, USA}

\begin{abstract}

Understanding the dynamics of the quiet solar corona is important for answering key questions including the coronal heating problem. Multiple studies have suggested small-scale magnetic reconnection events may play a crucial role. These reconnection events are expected to involve acceleration of electrons to suprathermal energies, which can then produce nonthermal observational signatures. However, due to the paucity of sensitive high-fidelity observations capable of probing these nonthermal signatures, most studies were unable to quantify their nonthermal nature. Here we use joint radio observations from the Very Large Array (VLA) and the Expanded Owens Valley Solar Array (EOVSA) to detect transient emissions from the quiet solar corona in the microwave (GHz) domain. While similar transients have been reported in the past, their nonthermal nature could not be adequately quantified due to the unavailability of broadband observations. Using a much larger bandwidth available now with the VLA and EOVSA, in this study, we are able to quantify the nonthermal energy associated with two of these transients. We find that the total nonthermal energy associated with some of these transients can be comparable to or even larger than the total thermal energy of a nanoflare, which underpins the importance of nonthermal energy in the total coronal energy budget.

\end{abstract}

\section{Introduction} \label{sec:intro}

The solar corona is generally classified into active regions and quiet solar corona. Active regions are defined as those regions of the solar corona which harbor strong magnetic fields and are bright in X-rays and extreme ultraviolet bands. They are responsible for large explosive events like solar flares, coronal mass ejections, etc. On the other hand, the quiet solar corona has weak magnetic fields and does not show such explosive behavior. However, with the development of each new generation of instruments, it has become increasingly clear that the quiet sun also shows a plethora of dynamics. The most recent example of this dynamics is the observation of ``campfires" \citep{berghmans2021} by the Extreme Ultraviolet Imager \citep{rochus2020} onboard the Solar Orbiter \citep{muller2020}. Quiet sun dynamics has also been reported by other authors \citep[e.g.,][]{krucker1997, kuhar2018, innes2001, harrison1999, berghmans1998, parnell2000, chitta2021,mandal2021,chen2019, voort2016, joshi2020, phillips2000}. We refer readers to reviews by \citet{shibasaki2011,madjarska2019,nindos2022} for a more detailed discussion on these topics. Observations of such extensive dynamics in the quiet sun are increasingly making the name ``quiet" a misnomer. 

There are also theoretical reasons as to why we expect to have a dynamic quiet solar corona as well. The most important of these reasons probably comes from the efforts made so far to understand the coronal heating problem \citep[see][for a review]{klimchuk2015}. All of the notable theories which try to explain this process suggest some dynamics in the quiet solar corona, although the exact dynamics can vary from theory to theory. 
For example, the nanoflare hypothesis \citep{Parker1988} suggests that a large number of nanoflares are continuously happening throughout the quiet corona, which in turn are responsible for maintaining the million-degree coronal temperature. This scenario requires the presence of ubiquitous small-scale magnetic-reconnection-driven nanoflare events throughout the corona. During the process of magnetic reconnection, particles get accelerated, although the extent to which this energization happens depends intimately on the details of the reconnection process \citep[see, e.g.,][]{bakke2018,james2018,glesener2020,frogner2020}. These accelerated particles can either escape into the interplanetary space and contribute to the suprathermal particle population observed in the solar wind \citep{wang2016,hou2021,mitchell2020,hill2020} or can get thermalized in the solar atmosphere and contribute to the energy budget of the quiet solar corona \citep[e.g.][etc.]{pontieu2011, frogner2020}. However, due to the rarity of studies regarding the energetics of these accelerated nonthermal electrons from the small microflares and nanoflares, their exact contribution to coronal heating is, as of now, unknown. 
Thanks to the availability of new instruments with high angular resolution, the recent detection of small-scale reconnection-driven spicules \citep{samanta2019}, ``campfires" \citep{berghmans2021}, and the suggestions that they might be related to the hypothesized nanoflares also make studies regarding detection of nonthermal electrons particularly interesting and timely.

While the presence of nonthermal electrons in the quiescent corona has been inferred using observations at the (extreme) ultraviolet wavelengths \citep[e.g.,][]{testa2014}, a more direct means of detecting these nonthermal electrons is through their emission at X-ray and radio wavelengths. Detection of X-ray transients from the quiet solar corona has already been reported \citep[e.g.][etc.]{kuhar2018,vadawale2021,paterson2022}. Confirmed detection of nonthermal electrons has also been reported for weak microflares with the Reuven Ramaty High Energy Solar Spectroscopic Imager and Nuclear Spectroscopic Telescope Array (NuSTAR) X-ray telescope thanks to its high sensitivity \citep[][]{qiu2004,kundu2006,stoiser2007, hannah2008,glesener2020,cooper2021}. 
Radio observations have also presented strong evidence regarding the presence of nonthermal electrons in the quiet solar corona. \citet{mondal2020, sharma2022, mondal2023} reported the detection of Weak Impulsive Narrowband Quiet Sun Emissions (WINQSEs) which the authors hypothesize are weaker cousins of the well-known type III radio burst and originate due to coherent plasma emission from nonthermal electrons generated in nanoflares \citep{mondal2021}. Recent studies at millimeter wavelengths have also revealed the presence of small-scale transients in the quiescent solar chromosphere \citep[e.g.][etc.]{nindos2020,eklund2020,nindos2021}. While these chromospheric transients generally occur due to thermal free-free emission, they can be driven by the precipitation of nonthermal electrons produced due to weak energy release events. For producing observable nonthermal gyrosynchrotron emission in these frequencies, a sufficiently large electron population with energy in the MeV range is required, which is not expected in the case of small-scale flares \citep{white1992}. 

Studies of these weak radio transients from the transition region and the low corona provide mixed results. While most studies suggest that these transients are free-free in nature \citep[see][for a review]{madjarska2019}, some suggest that some fraction of them are nonthermal in origin \citep{krucker1997, kundu2006}, implicating the acceleration of nonthermal electrons in small-scale energy release events. However, these studies were done a few decades back when broadband radio imaging spectroscopy observations were not available. This posed challenges in identifying the emission mechanism of these sources. Additionally, due to the weak nature of these sources, image fidelity has always been an issue. Here we use simultaneous detection of such sources with two instruments or over wide bandwidths to increase the fidelity of the detections.

In this paper, we use simultaneous broadband radio imaging spectroscopy data from the Very Large Array (VLA; \citealt{Perley2011}) and the expanded Owens Valley Solar Array (EOVSA; \citealt{Gary2018}) to detect weak radio transient sources and also obtain their spectral and temporal behavior when possible. We supplement these data with that obtained from the Atmospheric Imaging Assembly (AIA; \citealt{lemen2012}) onboard the Solar Dynamics Observatory (SDO; \citealt{pesnell2012}) to search for counterparts of these sources in the Extreme Ultraviolet (EUV) and use the EUV properties of these sources to delve deeper in their nature than that had been possible earlier.

This paper is structured as follows. In Section 2, we briefly describe the observation. Section 3 describes the details of the data analysis procedure. In Section 4 we present the results obtained, which are then discussed in a broader context in Section 5. Finally, we conclude with a brief summary in Section 6.

\section{Observations and context}

\begin{figure}
    \centering
    \includegraphics[trim={0.2cm 0.5cm 0 2cm},clip,scale=0.35]{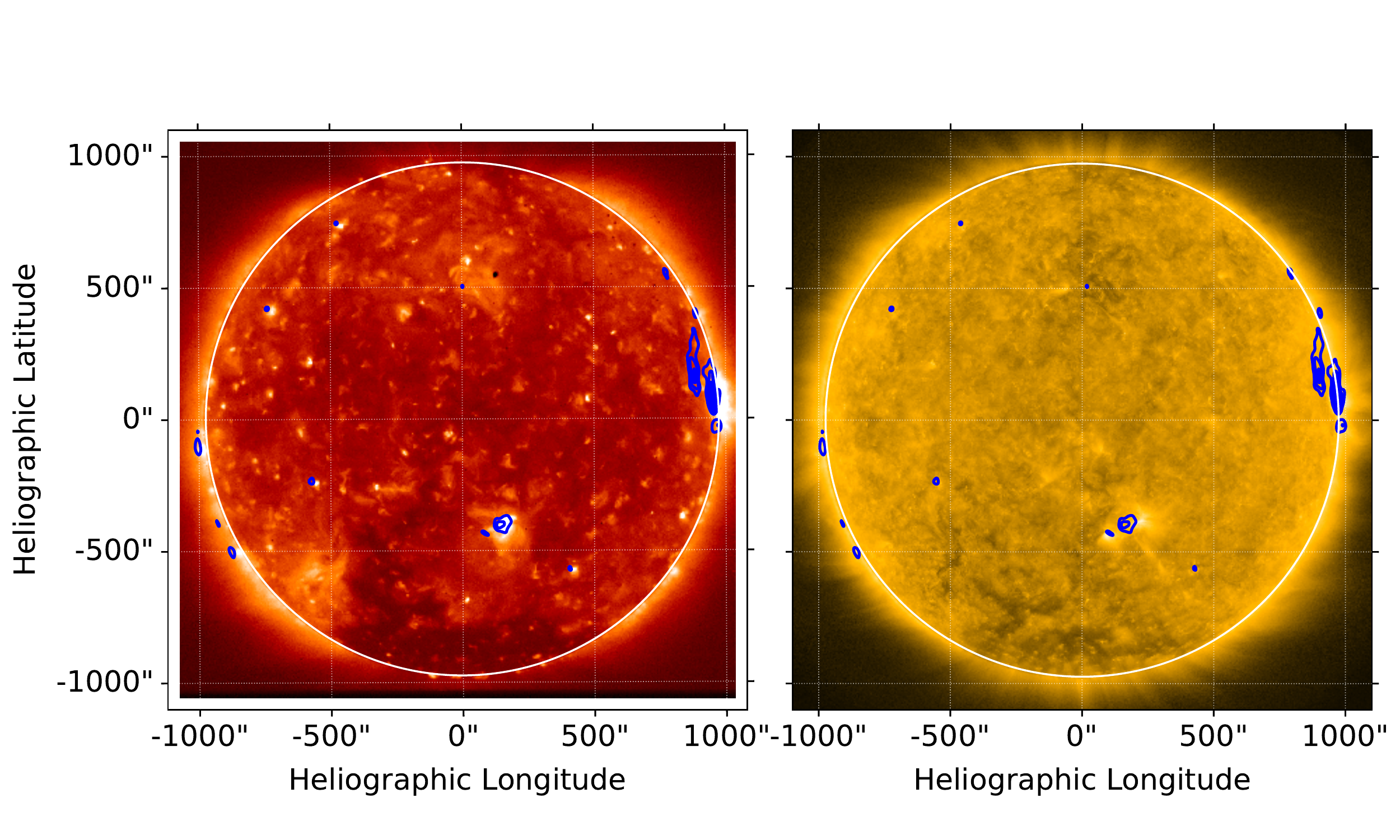}
    \caption{The full day integrated EOVSA 3 GHz radio map overlaid on the Hinode/XRT soft X-ray map (left panel) and the SDO/AIA 171\AA$\,$ EUV map (right panel).}
    \label{fig:context}
\end{figure}

The data presented here were acquired using the EOVSA and the VLA on February 1, 2020, between 19:00:00--23:00:00 UT. The Sun was extremely quiet on this day. No X-ray flare was reported by the Geostationary Operational Environmental Satellite (GOES). There was only one active region designated as National Oceanic and Atmospheric Administration (NOAA) 12757 presenting on the west limb. In addition, a bi-polar, but spotless magnetic structure is located on the disk (at S10W10). Fig. \ref{fig:context} shows contours of the full-day integrated radio map from the EOVSA\footnote{\url{http://ovsa.njit.edu/browser/?suntoday_date=2020-02-01}} at 3 GHz over an image from the Solar X-ray Telescope \citep[XRT;][]{deluca2000} onboard Hinode with the Al-poly filter (left panel) and an SDO/AIA 171 \AA$\,$ image (right panel). A radio counterpart is detected from each region against the quiet Sun disk. 

The VLA observed the Sun during this time in the subarray mode in its C-configuration, with one subarray operating in the P band and another operating in the L band (0.994--2.006 GHz). Here we present data from the L-band. {The entire band is divided into 8 spectral windows, each with a bandwidth of 128 MHz. Each spectral window has 128 channels and each channel has a width of 1 MHz. The maximum baseline length is 3.2 km, corresponding to an angular resolution of 14$''$--28$''$ in 1--2 GHz.} We have also analyzed EOVSA data from this time period, focusing primarily on the band spanning 1.425--1.749 GHz, which has a frequency closest to the analyzed VLA L band data.

\section{Data analysis}

\subsection{Calibration, imaging and source detection}

The VLA observation on this day did not have a flux calibrator observation. Hence, approximate flux densities were estimated assuming that the flux of the phase calibrator, J1941-1524, is the same as that given in the VLA Calibrator manual\footnote{\url{http://www.vla.nrao.edu/astro/calib/manual/csource.html}} at all frequencies. The flux density at 20 cm and 6 cm were used to determine the spectral index, which was then used to obtain the frequency-dependent flux density at each spectral window. All analysis was done using the Common Astronomy Software Applications \citep[CASA,][]{mcmullin2007}. 
At the time of the observation, spectral windows 1 (centered at 1.185 GHz) and 4 (centered at 1.557 GHz) were heavily affected by radio frequency inference (RFI) and were excluded from further analysis. {Details of the calibration and data analysis procedure are provided in the Appendix.} 
Calibrated EOVSA data were obtained from the observatory for this day following the standard calibration procedure.

Due to the large volume of data, automated techniques were used both for deconvolution and source identification purposes. For imaging the VLA data, we divided the entire duration into 15-minute chunks and imaged each chunk using the image deconvolution task \texttt{tclean} with the auto-masking mode. Unless specified otherwise, the time integration and frequency integration is 15 minutes and 128 MHz, respectively. {Details of the imaging procedure are provided in the Appendix.}
However, we find this technique is not very useful for processing the EOVSA data for the weak radio transients due to the particularly high sidelobe levels ($\sim 70$\%). Hence the EOVSA data were imaged manually and sources were identified with visual inspection.

\subsection{Filtering scheme to remove probable spurious sources}

Careful considerations were taken to ensure that no spurious source is included in the imaging results used for further analysis. The automasking algorithm of \texttt{tclean} which was used to identify real sources during the deconvolution procedure was tuned such that it only identifies sources that are higher than the 1.5 times the maximum sidelobe level of the brightest sources identified earlier. During the image deconvolution, frequent major cycles \citep{schwab1984}\footnote{The CLEAN algorithm has two major steps, namely the minor cycle and major cycle. During the minor cycle, model sky components are found in the image plane iteratively. After each model component is found, a scaled and truncated copy of PSF centered at that component is subtracted. During the major cycle, all the found components are Fourier transformed to the visibility plane and then subtracted from the observed visibilities. For more details, we refer the reader to \url{https://science.nrao.edu/science/meetings/2018/16th-synthesis-imaging-workshop/talks/Wilner_Imaging.pdf}.} were performed to ensure that the artifacts due to inaccurate subtraction of the instrumental point spread function are minimal.
However, in the spirit of additional caution, we considered the possibility that still some sources might be sidelobes of real sources. The possibility of spurious source detection is much more in the EOVSA data. The sidelobe levels are very high ($\sim 70\%$ at $\sim$1$'$ away from the main beam at 1.5 GHz) and hence it often becomes impossible to determine the true location of a source when multiple closely spaced sources are present in the image. While care was taken to minimize these effects, we found that the best way to find true sources is to compare the images obtained using data having very different uv-distributions and PSFs. Both the uv-distribution and image artifacts change significantly between images with very different frequencies. Hence if a source is detected at different frequencies, it is considered to be a real source. If a source is detected at a single frequency in the VLA/EOVSA, we investigate if the source is also detected by the other instrument. For additional caution, we also compare the flux density of the source in both images if it is detected by both instruments. Due to the difference between the frequency and flux scale of the two instruments we assume a 15\% and 10\% systematic flux uncertainty in the case of EOVSA and VLA, respectively. If under these assumptions the flux and location of the source match in the EOVSA and VLA, it is considered to be a real source. Additionally, the brightest source in each image is also believed to be real as it is highly unlikely that other real sources in the image conspire such that their sidelobes merge together to produce the brightest source in the image. To summarize a source is considered to be real if it satisfies at least one of the following criteria:

\begin{enumerate}
    \item It is the brightest source in the deconvolved image.
    \item The same source is detected at multiple frequencies over a wide bandwidth.
    \item The same source is detected in both EOVSA and VLA deconvolved images and their flux densities are consistent within uncertainties.
\end{enumerate}

\section{Results}

Using the techniques described in the previous section, we have detected several sources with high confidence, some lying on coronal holes, some associated with coronal bright points, some located beyond the limb, and some lying above the network regions. Below we discuss examples of these different types of sources. We have not included sources associated with active regions as these sources have been studied quite extensively and are not the focus of this work.

\subsection{Beyond the limb source}\label{sec:limb}

\begin{figure*}
    \centering
    \includegraphics[scale=0.4]{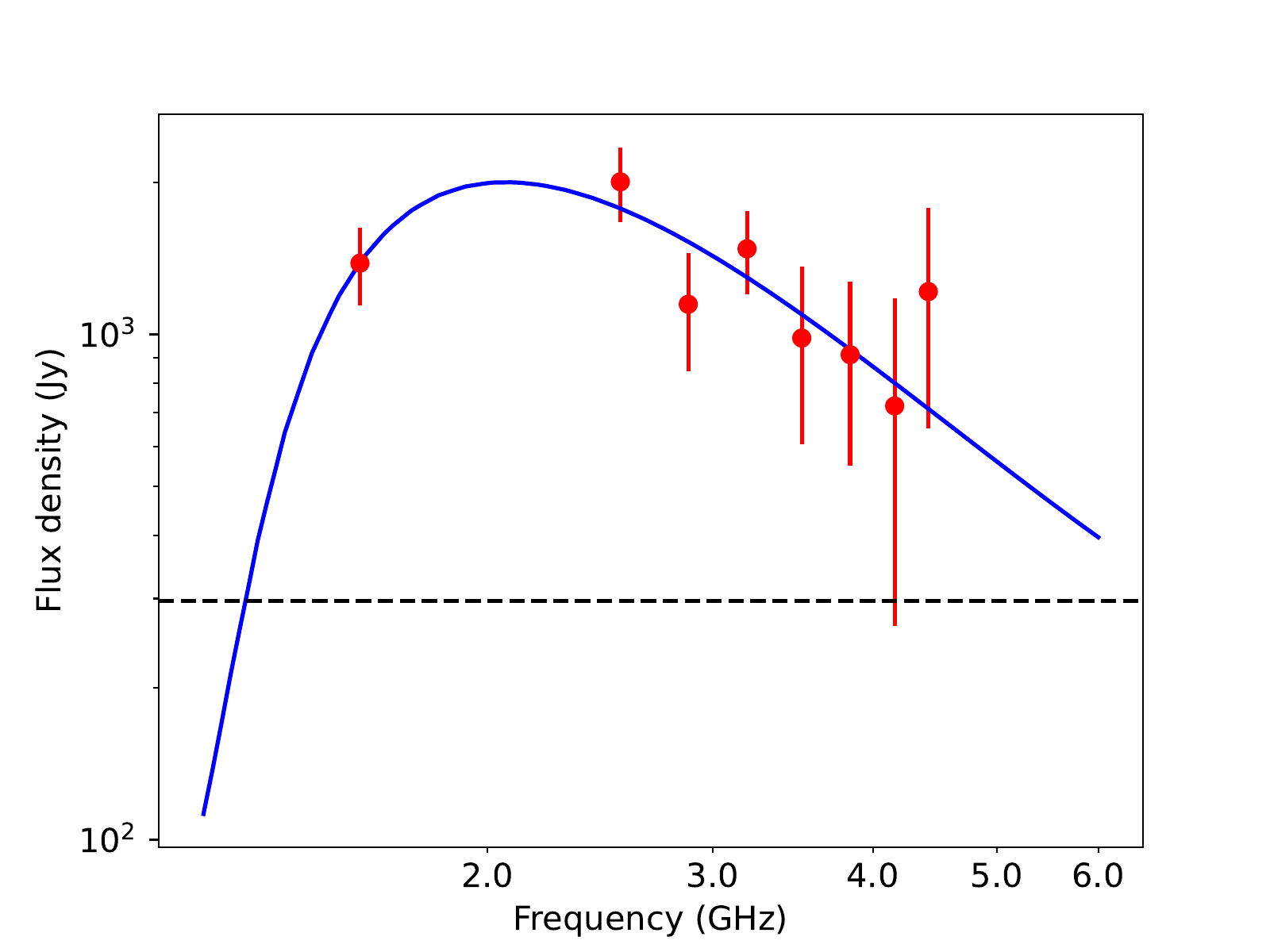}
    \includegraphics[trim={0 0 3cm 0},clip,scale=0.4]{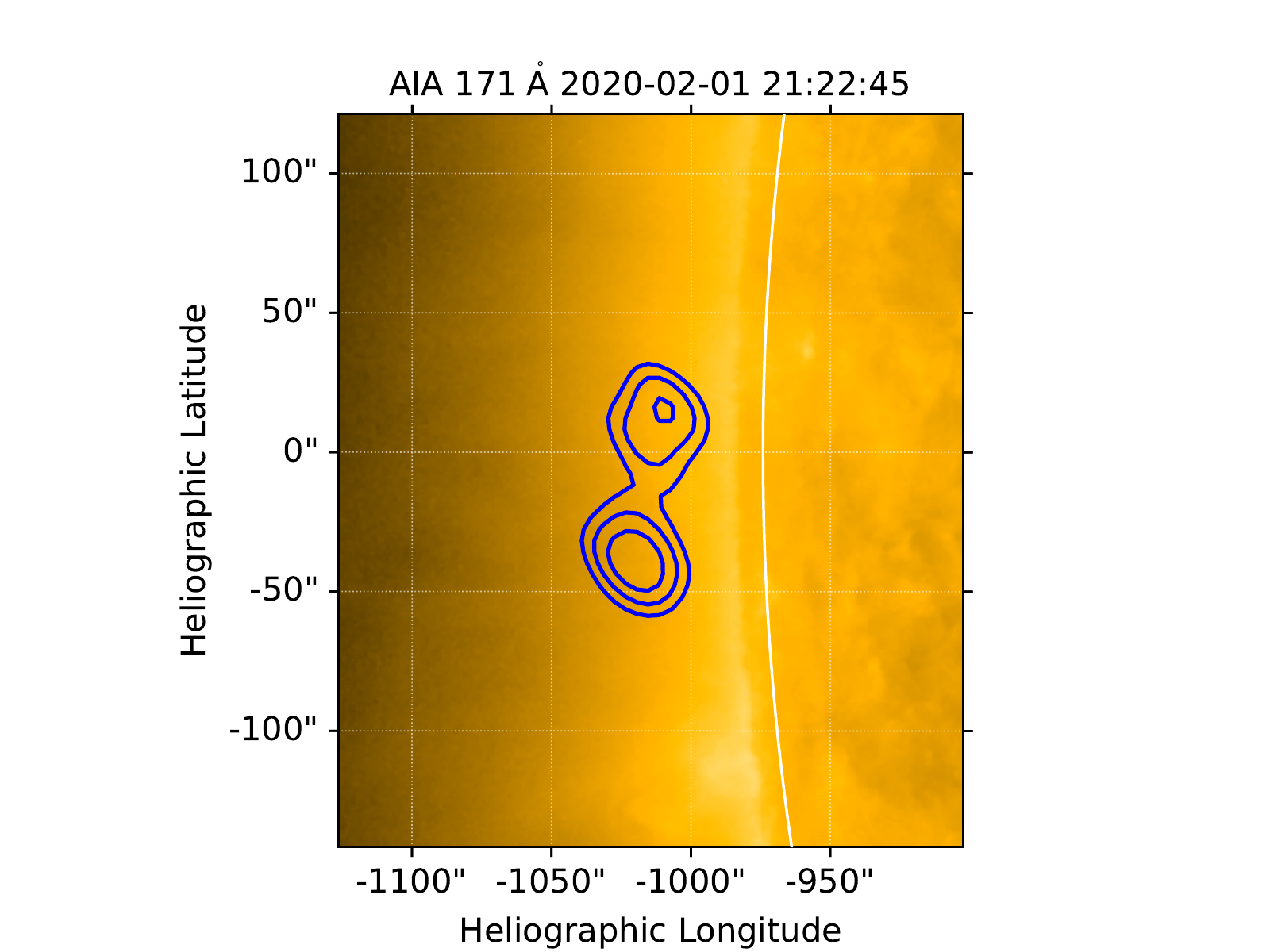}
    \caption{Left panel: The red points show the observed spectrum of the above-the-limb source (S1 in table \ref{tab:beyond_limb_params}) between 21:15--21:30.  The blue line shows the fitted gyrosynchrotron model. The black dashed line shows the expected spectrum if the emission due to optically thin free-free radiation. Right panel: Contours of the radio source at 1.6 GHz overlaid over a SDO/AIA 171\AA$\,$ image. }
    \label{fig:beyond_limb}
\end{figure*}

 While the integration time of EOVSA is 1 second, we have averaged the data over 15 minutes for this work. This was done in order to increase the image signal-to-noise ratio (SNR), and also for the ease of exploring the entire 4 hours of data with manual imaging. This source was first identified at 1.5 GHz using EOVSA data by averaging over 15 minutes between 21:15:00--21:30:00. It was also detected at a few other frequencies, although significant spatial averaging was needed for this. In the right panel of Fig. \ref{fig:beyond_limb} radio contours of this source are overlaid on a nearby AIA 171\AA$\,$ image. The spectrum of the source is shown in the left panel. The spectrum has been determined by taking the peak value of this source after smoothing each image to a resolution of  $150^{''}\times 150^{''}$. At this resolution, the source becomes unresolved, and hence the peak value (in units of Jy/beam) is equal to the integrated flux density (in units of Jy). {The error bars were obtained by adding the rms of the image and a 15\% flux uncertainty in quadrature. }
We have done a differential emission measure analysis using publicly available code\footnote{\url{https://github.com/ianan/demreg/tree/master/python}} following \citet{hannah2012,hannah2013} and use the output of that to calculate the expected free-free emission using the code developed in \citet{fleishman2021}. The expected flux density due to free-free emission is shown using a black dashed line. {It has been calculated directly from the model solar emission, by summing the pixel values inside the outermost contour shown in the right panel of Fig. 2.} It is evident that the expected flux density is much lower than that observed in the EOVSA radio images. { However, it should be noted that there can be small differences because the EUV limb and the radio limb do not lie at the same location. Additionally, the EUV images do not seem to have a counterpart to the observed radio source. We have also simulated an EOVSA observation using the free-free emission model and produced simulated images using the same procedure as that followed while producing the observed EOVSA map. No structure, similar to that observed in the EOVSA map, is detected in the simulated map, even though the rms of both the simulated and observed maps are comparable. }
Based on this we rule out free-free emission as the dominant emission mechanism behind the observed source.

Next, we investigate if a gyrosynchrotron model can explain the observed spectrum. The gyrosynchrotron model was obtained using the code available in \citet{kuznetsov2021,fleishman2021}. The source is assumed to be homogeneous and isotropic.  
The nonthermal electron distribution is assumed to be a powerlaw between $E_{min}$ and $E_{max}$ with powerlaw index $\delta$. 
The magnetic field (B), $\delta$, and number density of nonthermal electrons were kept as free parameters during the fitting procedure. We also manually adjust other parameters to obtain a good fit to the data. The density of the thermal and nonthermal electrons are denoted as $n_{\rm th}$ and $n_{\rm  nth}$ respectively. The angle between the LOS and the magnetic field is denoted as $\theta$. The parameters used for modeling the spectra are given in Table \ref{tab:beyond_limb_params} in the top row, marked with the label S1. The fitted spectrum is shown in the left panel of Fig. \ref{fig:beyond_limb} using a blue line. { It appears that for this source, $n_{\rm nth}$ is greater than $n_{\rm th}$, which, while rather uncommon, has been reported earlier in some flare events \citep{Krucker2010,fleishman2016}. We have been unable to find a valid model in which $n_{\rm nth}$ is smaller than the assumed thermal density, even when the magnetic field is allowed to be as high as 5 kG. While we have provided the formal uncertainties for the fitted parameters, due to the insufficient frequency sampling and bandwidth coverage, the fit is highly under-constrained. The fit parameters given in the table are certainly non-unique and should only be taken as representative values. Thus the primary source of uncertainties in those fit parameters is largely due to the systematic uncertainties due to the under-constrained fitting.} 
Using these representative fit parameters, we calculate that the total nonthermal power of this source is $2.6\times 10^{26}\,$ergs s$^{-1}$. Assuming a constant power over the entire integration time, we estimated the total nonthermal energy associated with this source to be $2.3\times 10^{29}\,$ergs, which is comparable to that estimated in the case of microflares based on X-ray data \citep{hannah2008,hannah2008b,glesener2020}. 

\begin{table*}
\centering
\begin{tabular}{|p{1.2cm}|p{1.5cm}|c|p{1.5cm}|p{1cm}|p{1.5cm}|p{1.5cm}|p{1cm}|p{1.5cm}|p{1.5cm}|p{1cm}|}
\hline \hline

& Magnetic field (G) & $\delta$ & E$_{max}$ (MeV) & $\theta$ (deg) & $\log_{10}$($nth$ (cm$^{-3}$))& 
Temper- ature (MK) &  LOS depth (arcsec) & Area (arcsec$^2$) & $\log_{10}$($nnth$ (cm$^{-3}$) & E$_{min}$ (keV) \\ \hline

S1 & 47$\pm$7 & 6.4$\pm$0.3 & 10 & 50 & 9.5 & 3 & 6.41  & 326 & 9.8$\pm$0.2 & 30 \\ \hline
S2 & 129$\pm$12 & 5.7$\pm$0.2 & 10 & 70 & 9.4 & 3  & 5 & 100 & 9$\pm$0.7 & 1 \\ \hline
  
 \end{tabular}
 \caption{Parameters used to obtain the model gyrosynchrotron spectrum shown in Figs. \ref{fig:beyond_limb} and \ref{fig:vla_source1_spectrum}. { Note due to the under-constrained nature of the spectral fitting, they should only be regarded as representative values. }}
 \label{tab:beyond_limb_params}
 \end{table*}

\subsection{Source associated with coronal bright point} \label{sec:cbp}

\begin{figure*}
    \centering
    \includegraphics[scale=0.5]{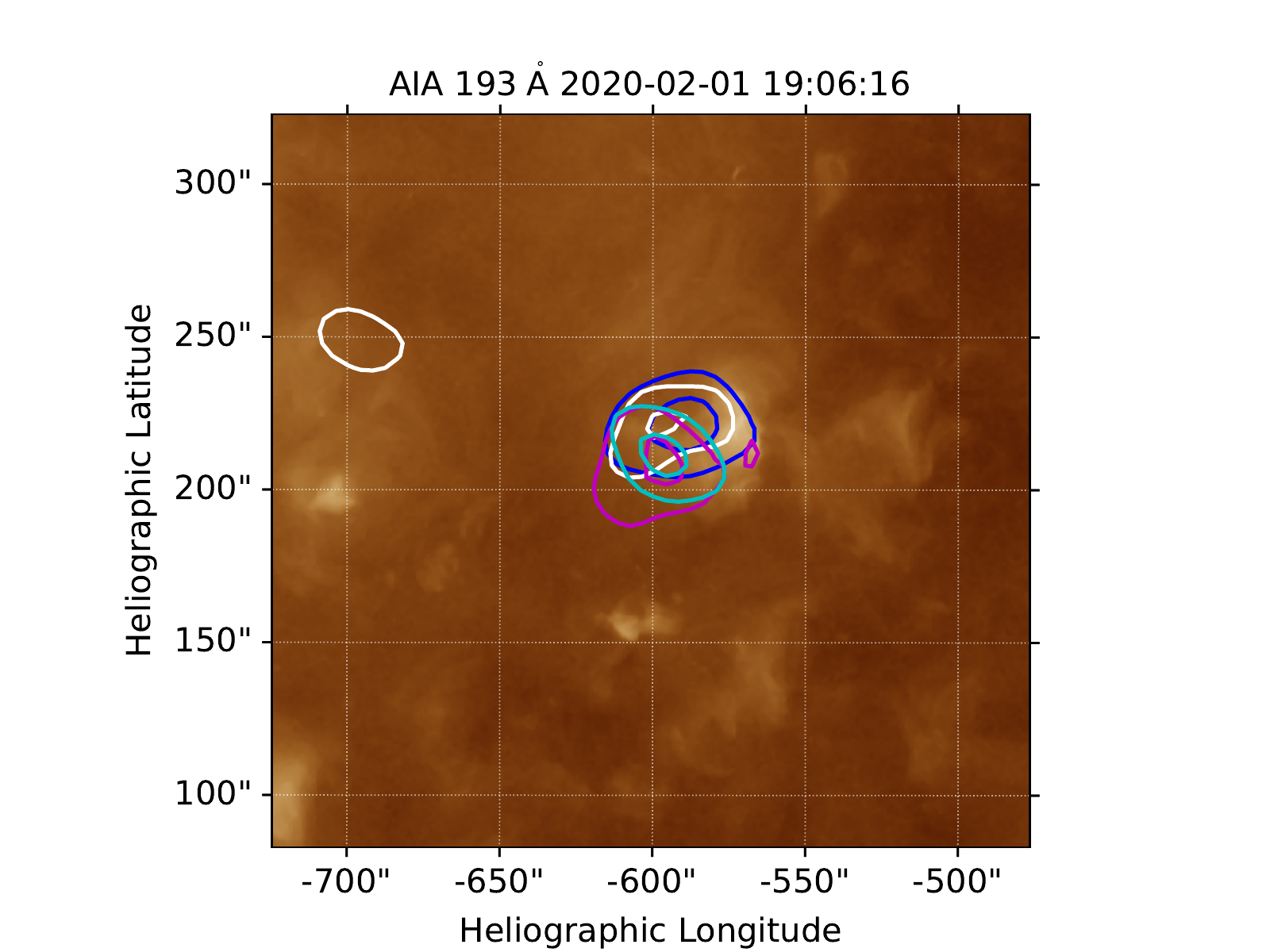}
    \includegraphics[scale=0.5]{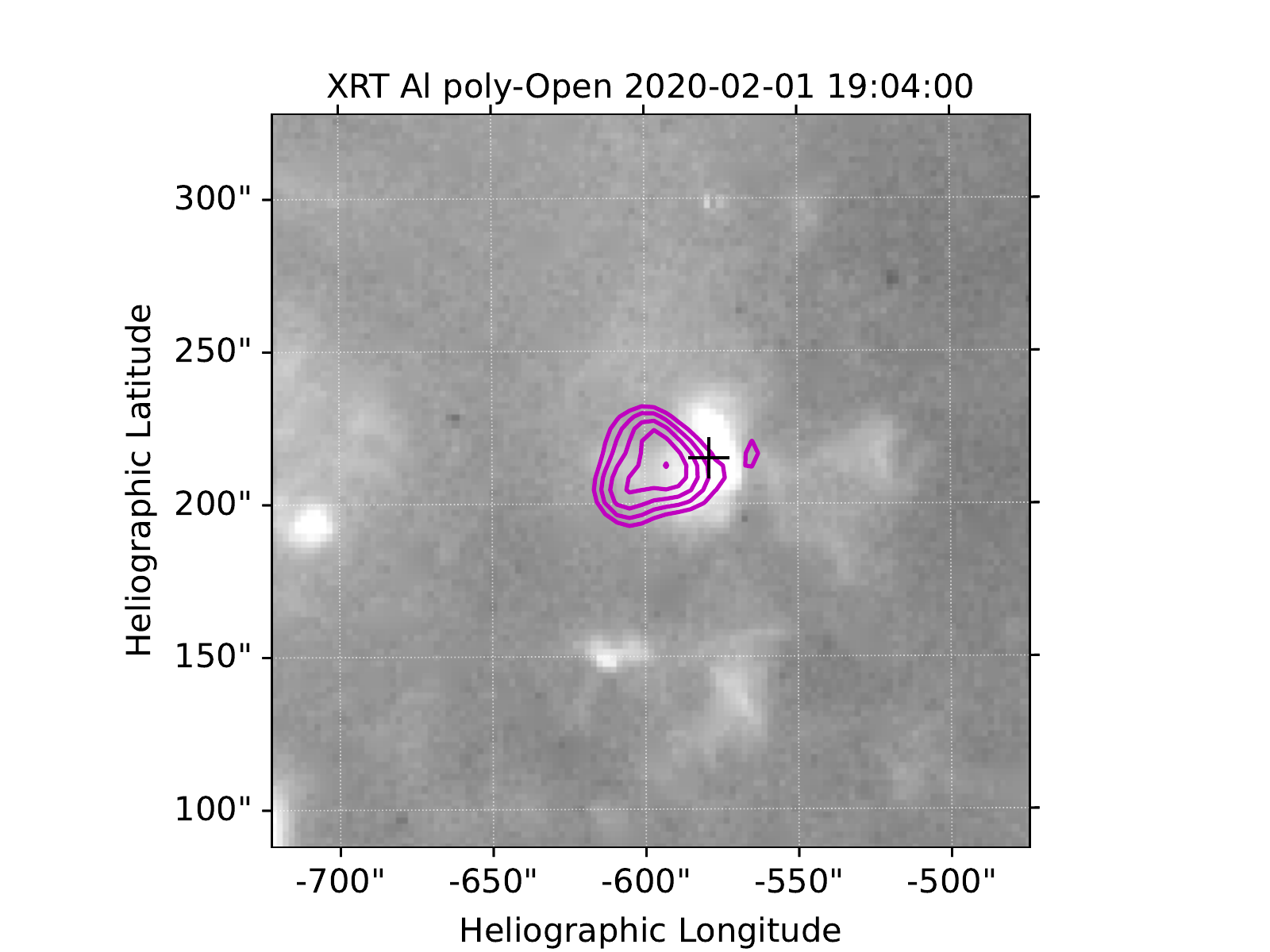}
    \caption{Radio contours of the source (S2 in table \ref{tab:beyond_limb_params}) described in Sec \ref{sec:cbp} are overlaid on a AIA 193\AA$\,$ image (left panel) and a XRT Al-poly image (right panel). The blue, white, magenta and cyan contours correspond to images made at 1.057, 1.313, 1.441 and 1.685 GHz respectively.}
    \label{fig:coronal_bright_point_source}
\end{figure*}

\begin{figure}
    \centering
    \includegraphics[scale=0.5]{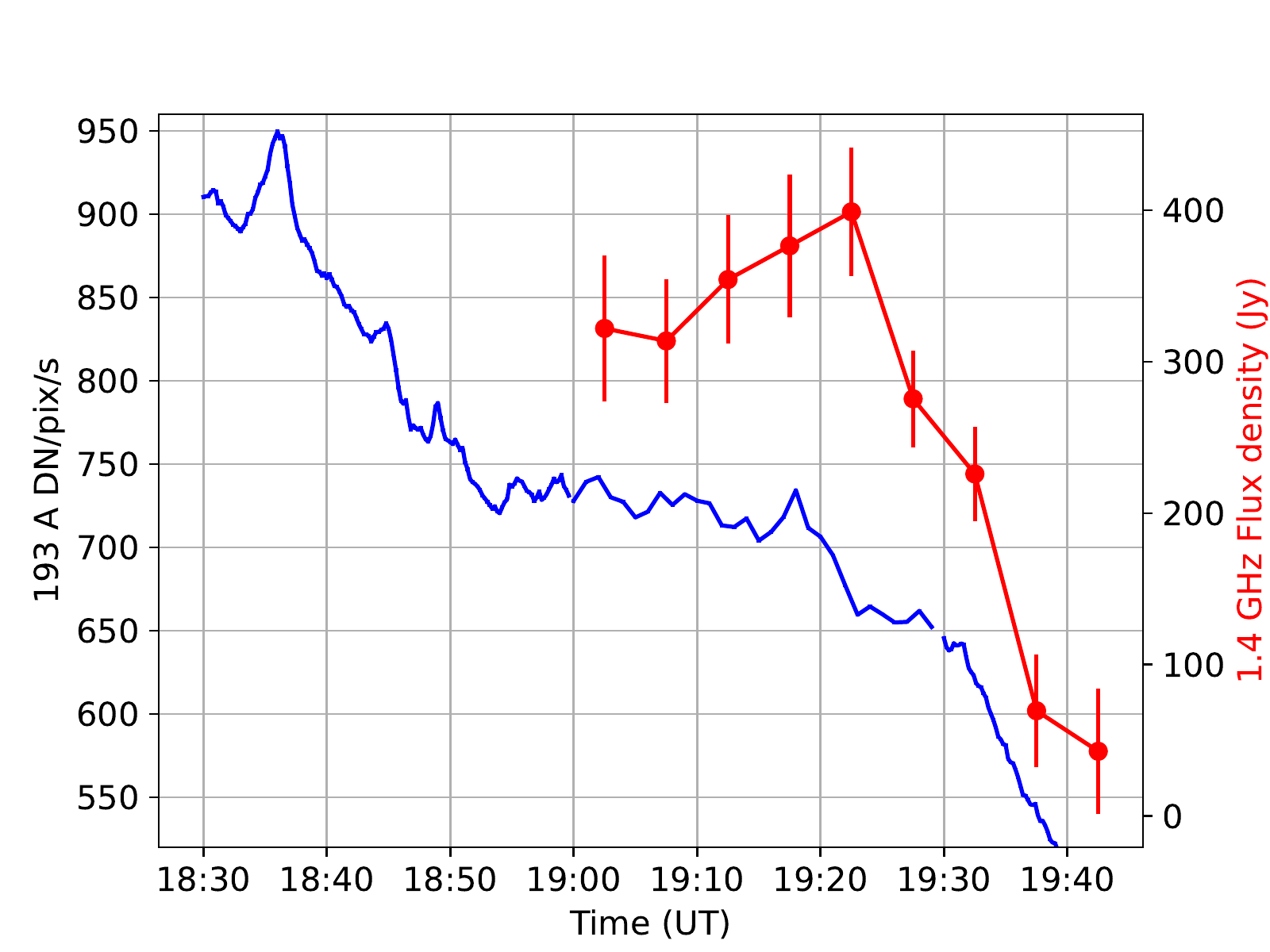}
    \caption{Light-curve at 193\AA$\,$ from a 24$^"\times$14$^{"}$ box centered on the black cross shown in Fig. \ref{fig:coronal_bright_point_source} is shown in blue. The 1.4 GHz light curve is shown in red.}
    \label{fig:vla_source1_aia_lightcurve}
\end{figure}

This source was detected at 4 frequencies in the VLA data between 19:00:00-19:15:00. Radio contours of this source (S2 in table \ref{tab:beyond_limb_params}) are overlaid on an SDO/AIA 193\AA$\,$ image at 19:06:16 in the left panel of Fig. \ref{fig:coronal_bright_point_source}. The blue, white, magenta, and cyan contours correspond to images made at 1.057, 1.313, 1.441, and 1.685 GHz respectively. This source also had a signature in the soft X-ray wavelengths, shown on the right, where the radio 1.441 GHz contour is overlaid on a Hinode/XRT Al-poly image. This source was only detected at 1.4 GHz in the later time period 19:15:00--19:30:00 and is not detected at any frequency at later times. {In Fig. \ref{fig:vla_source1_aia_lightcurve}, we show the 193 \AA$\,$ and 1.4 GHz light curve. {The 1.4 GHz light curve was obtained by making images with a time integration of 5 minutes.} The coronal bright point (CBP) shows a clearly decreasing flux, which implies that it was in the decay phase during this time. Interestingly we find that during the time when radio data were available and the source was detected, the 1.4 GHz light curve of this source and the EUV light curve show very similar trends.}

\begin{figure}
    \centering
    \includegraphics[scale=0.5]{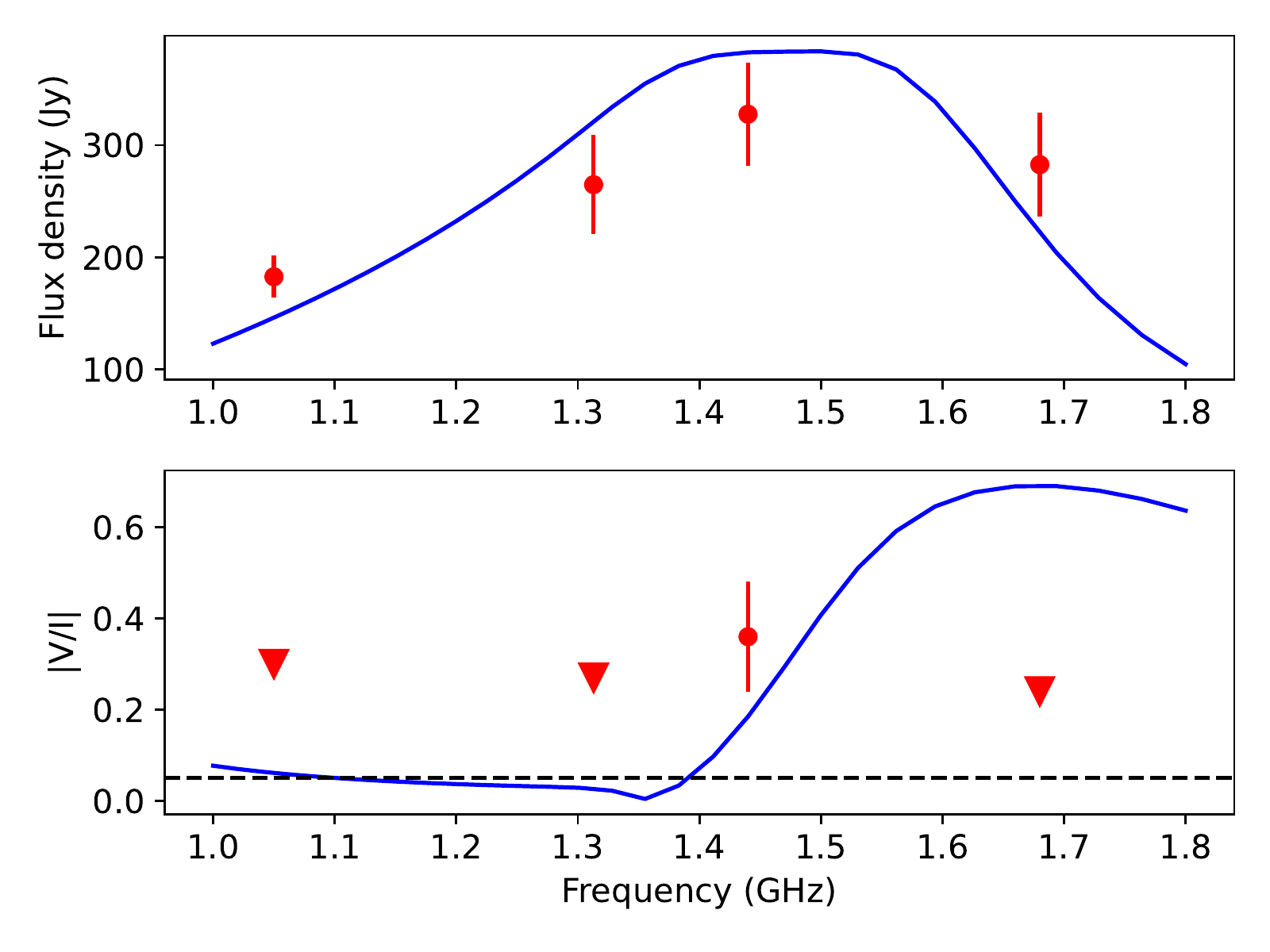}
    \caption{Red and blue indicate the observed values and the modeled gyrosynchrotron spectrum respectively for the source described in Sec \ref{sec:cbp}. The red triangles denote the upper limits obtained using the 3 times the rms in the Stokes V image. The black dashed line shows the expected value assuming free-free emission and a medium with a magnetic field of 100 G.}
    \label{fig:vla_source1_spectrum}
\end{figure}

In the upper panel of Fig. \ref{fig:vla_source1_spectrum}, the spectrum of this source is shown. The {absolute value} of the stokes V/I spectrum of this source is also shown in the bottom panel. {The error bars have been determined by adding a 10\% systematic flux uncertainty in quadrature to the rms of the image. The red triangles denote the upper limits obtained using the 3 times the error in the estimated V/I and are shown when stokes V is not detected with at least an SNR of 3.} The blue curve shows a model gyrosynchrotron spectrum which can account for the detected flux and the degree of circular polarization at different frequencies. {It is evident that the model predicts a higher degree of circular polarization at 1.68 GHz, than that expected from the upper limit at that frequency. Such behavior is possible if there are inhomogeneities in the source, leading to a decrease in the observed circular polarization. However, exploring the details of this is beyond the scope of this work.} The parameters used for modeling the spectra are given in Table \ref{tab:beyond_limb_params} in the top row, marked with the label S2. The nonthermal energy flux estimated using the model parameters is $7 \times 10^{22}\,$erg s$^{-1}$. Again assuming a constant power over the entire integration time, we estimate the nonthermal energy associated with this source as $6\times 10^{25}\,$ergs. The average thermal energy estimated from DEM analysis for this source is $5\times 10^{25}\,$ergs, comparable to the estimated nonthermal energy of the microwave source.

Past studies of CBPs have assumed that the radio emission associated with them is generated due to free-free emission \citep{habbal1986}. A high degree of circular polarization, if detected, was explained by propagation effects through a magnetized plasma. We investigate this possibility using simulated free-free emission maps, generated by the procedure described in Section \ref{sec:limb}. We assume that the medium has a uniform magnetic field of 100 G. The black dashed line shows the expected degree of circular polarization from this simulation.  It is evident that the simulated values are much lower than those observed. Hence we conclude that free-free emission cannot be the emission mechanism behind the observed source. 

\subsection{Sources associated with coronal holes} \label{sec:coronal_hole_sources}

\begin{figure*}
    \centering
    \includegraphics[scale=0.38]{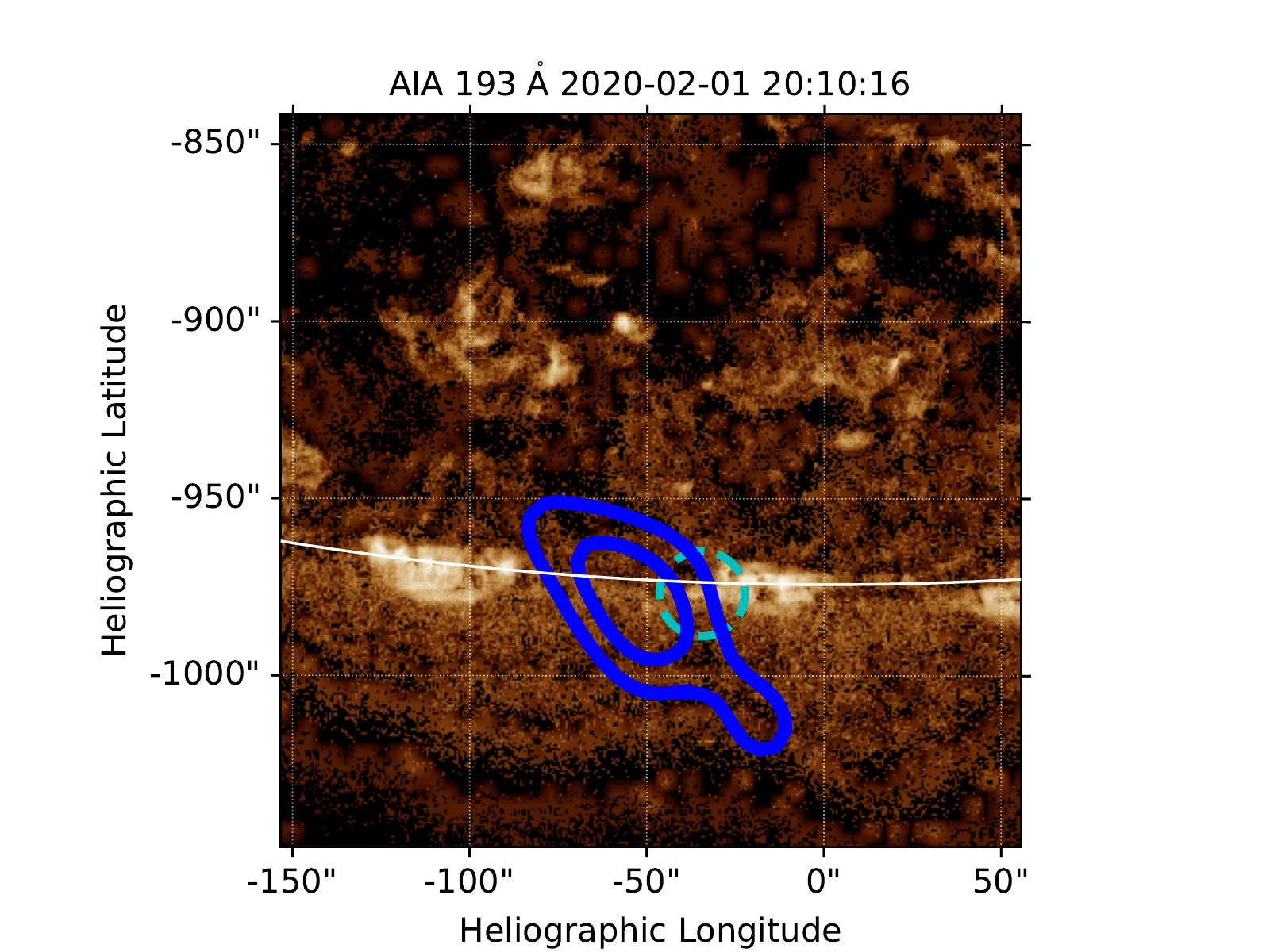}
    \includegraphics[trim={2cm 0 0 0},clip,scale=0.38]{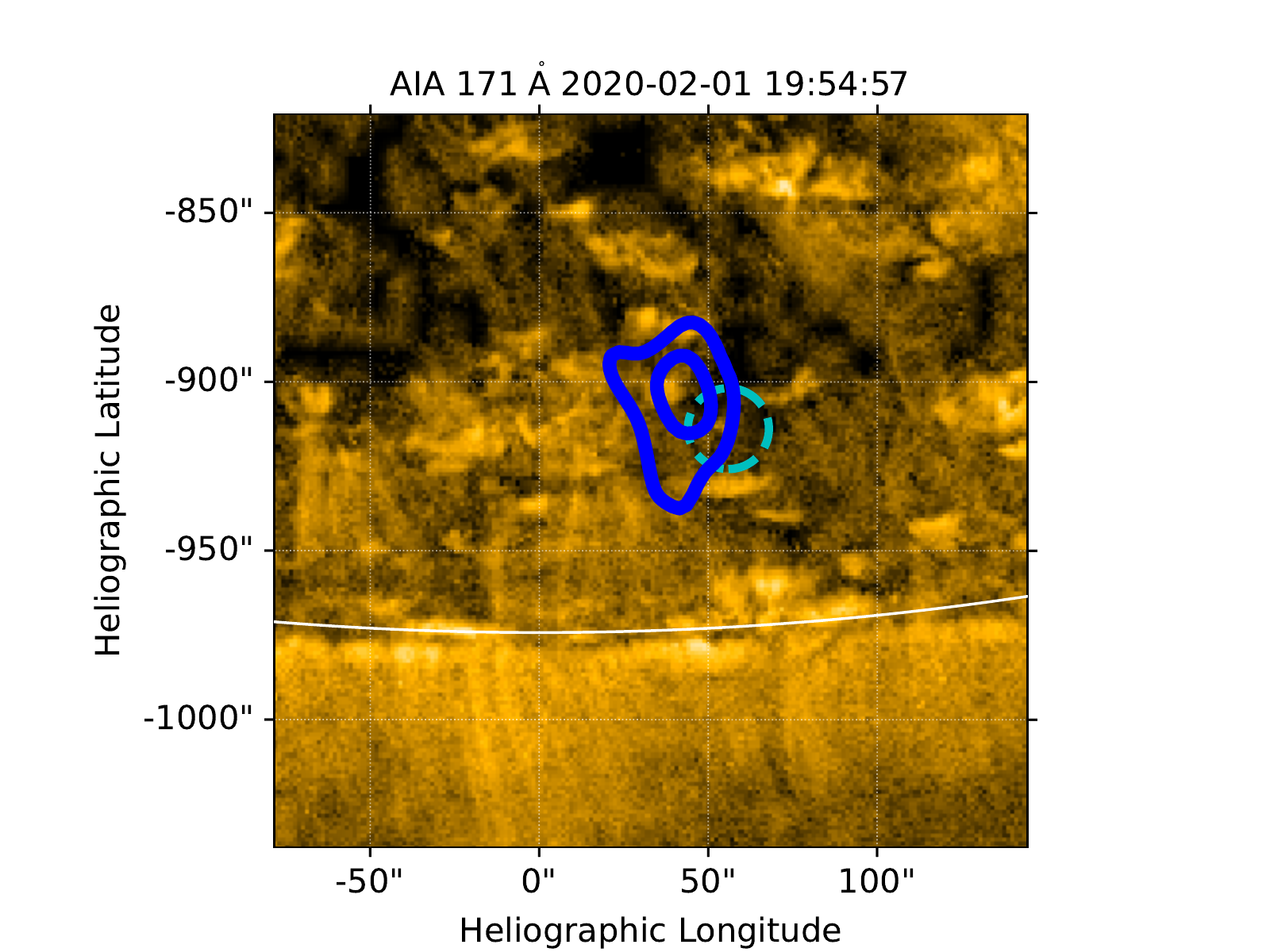}
    \includegraphics[trim={2cm 0 0 0},clip,scale=0.38]{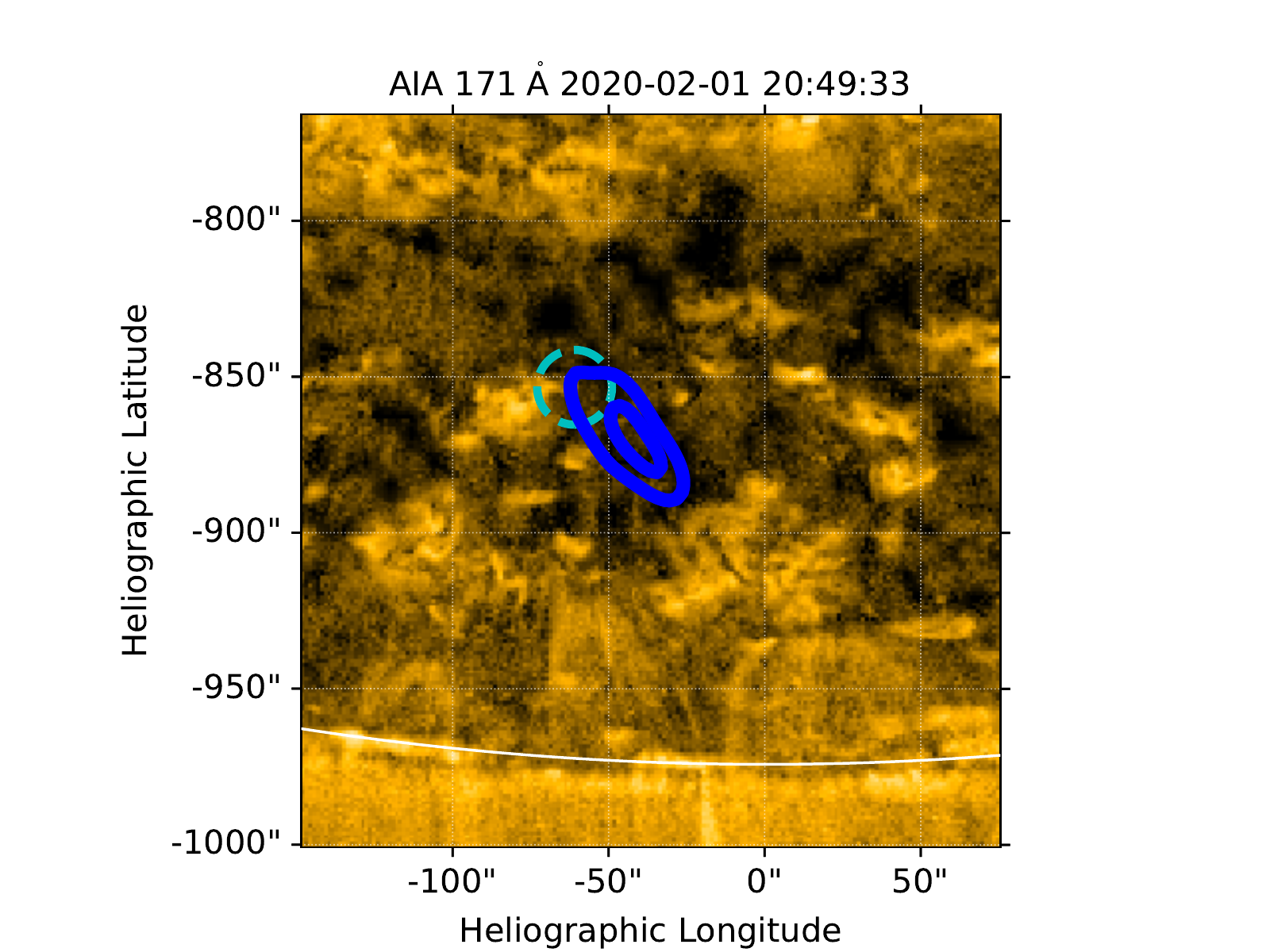}
     \includegraphics[scale=0.35]{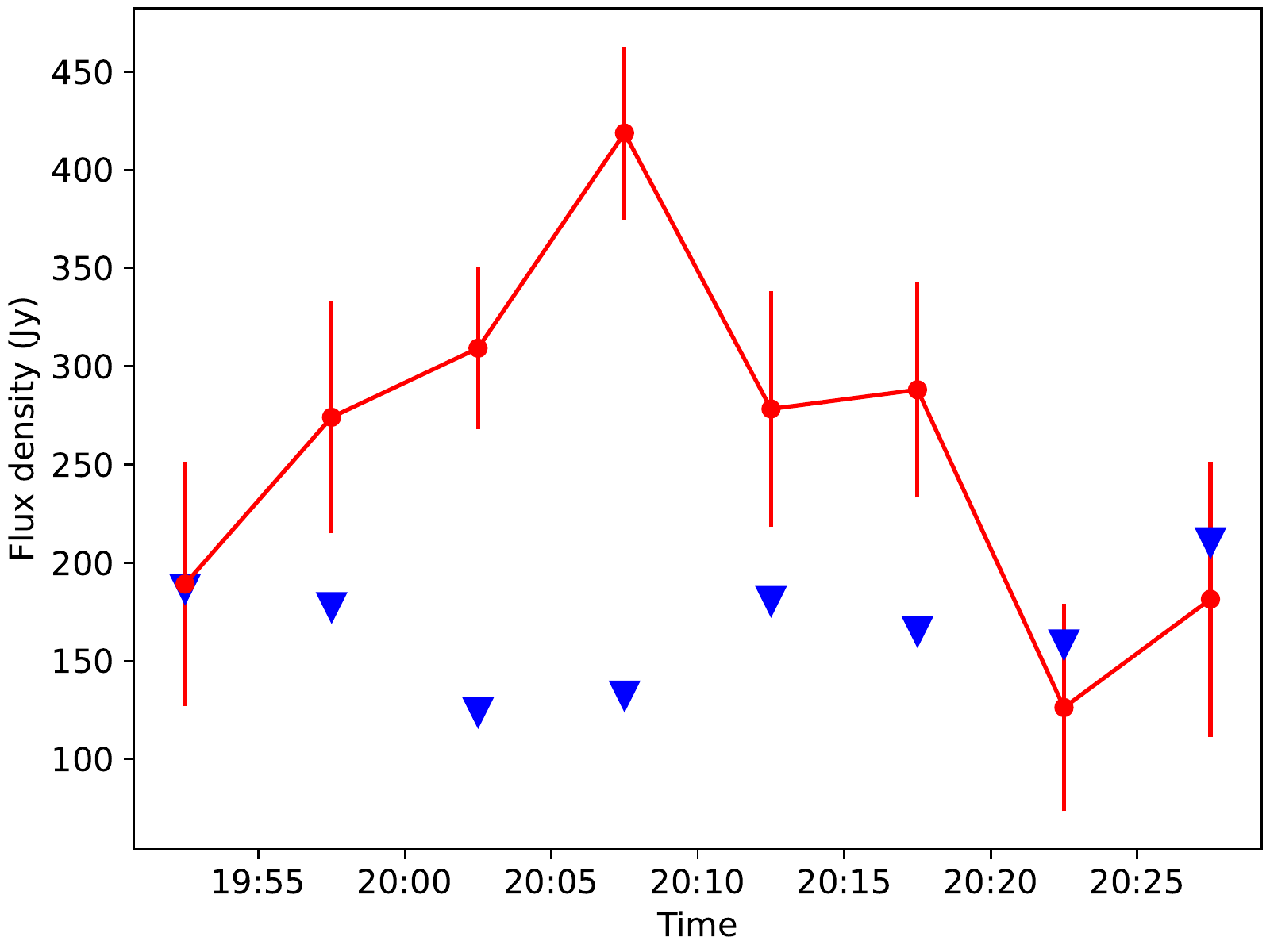}
      \includegraphics[trim={1cm 0 0 0},clip,scale=0.35]{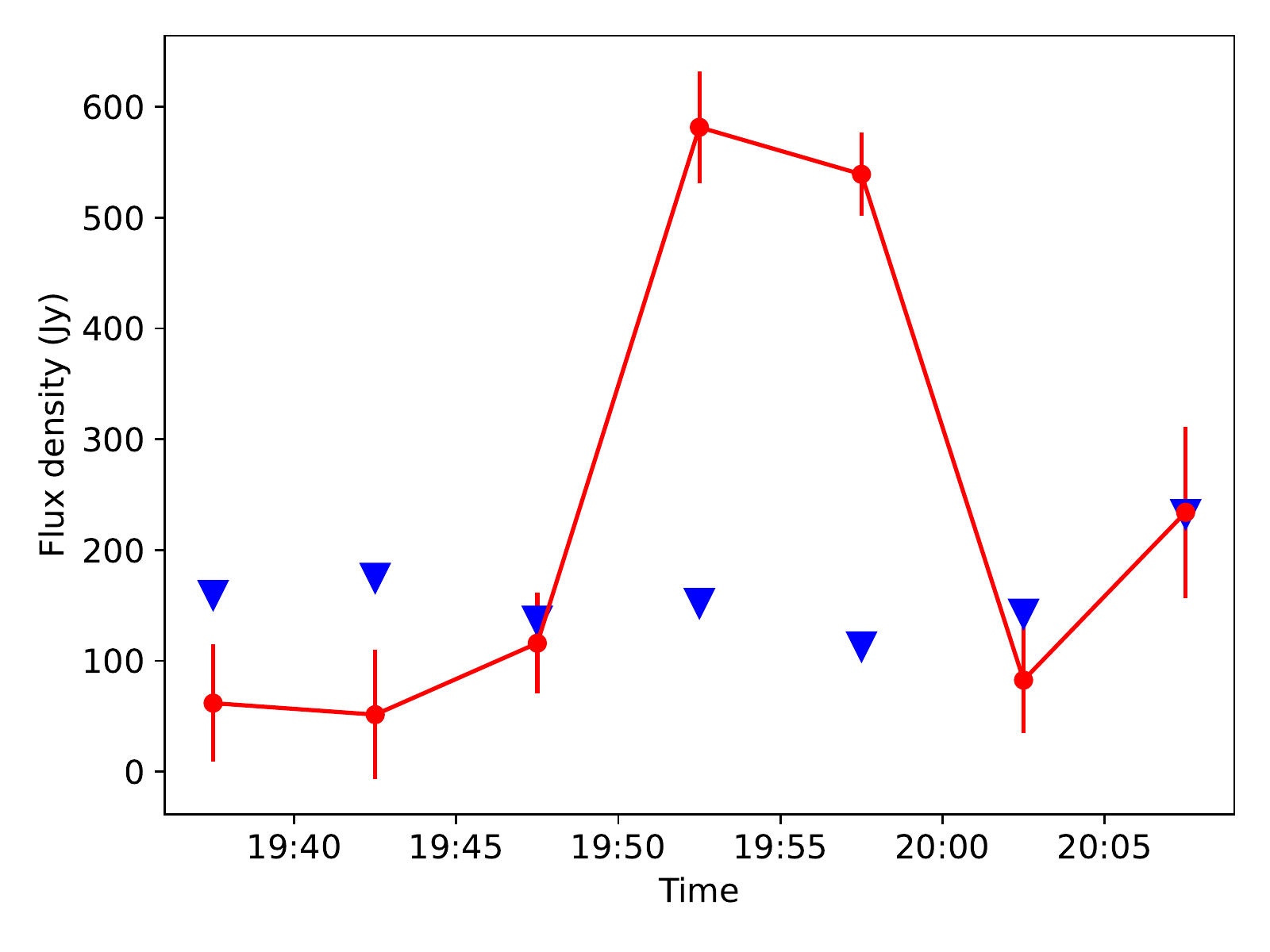}
       \includegraphics[trim={1cm 0 0 0},clip,scale=0.35]{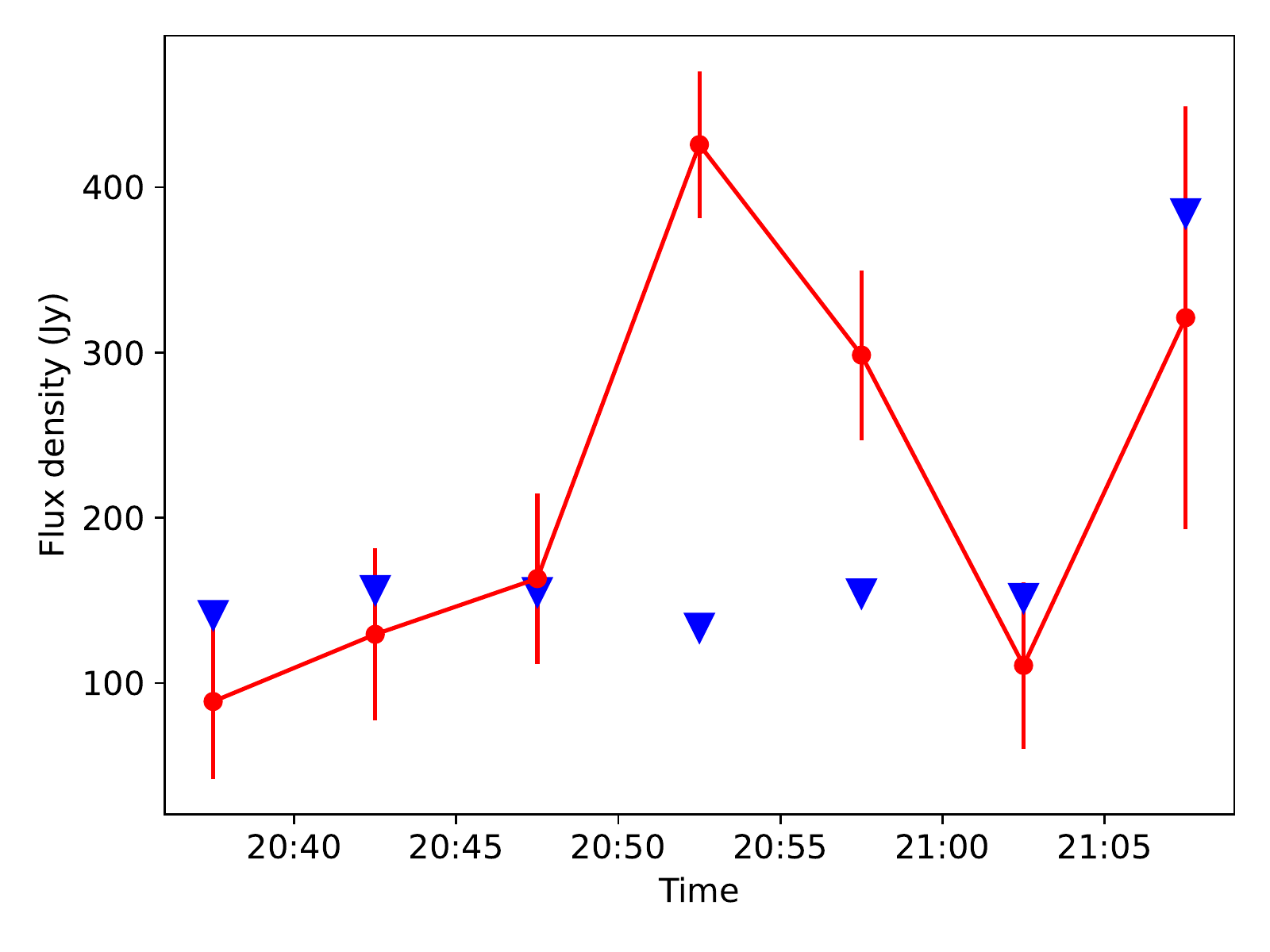}
       \includegraphics[scale=0.38]{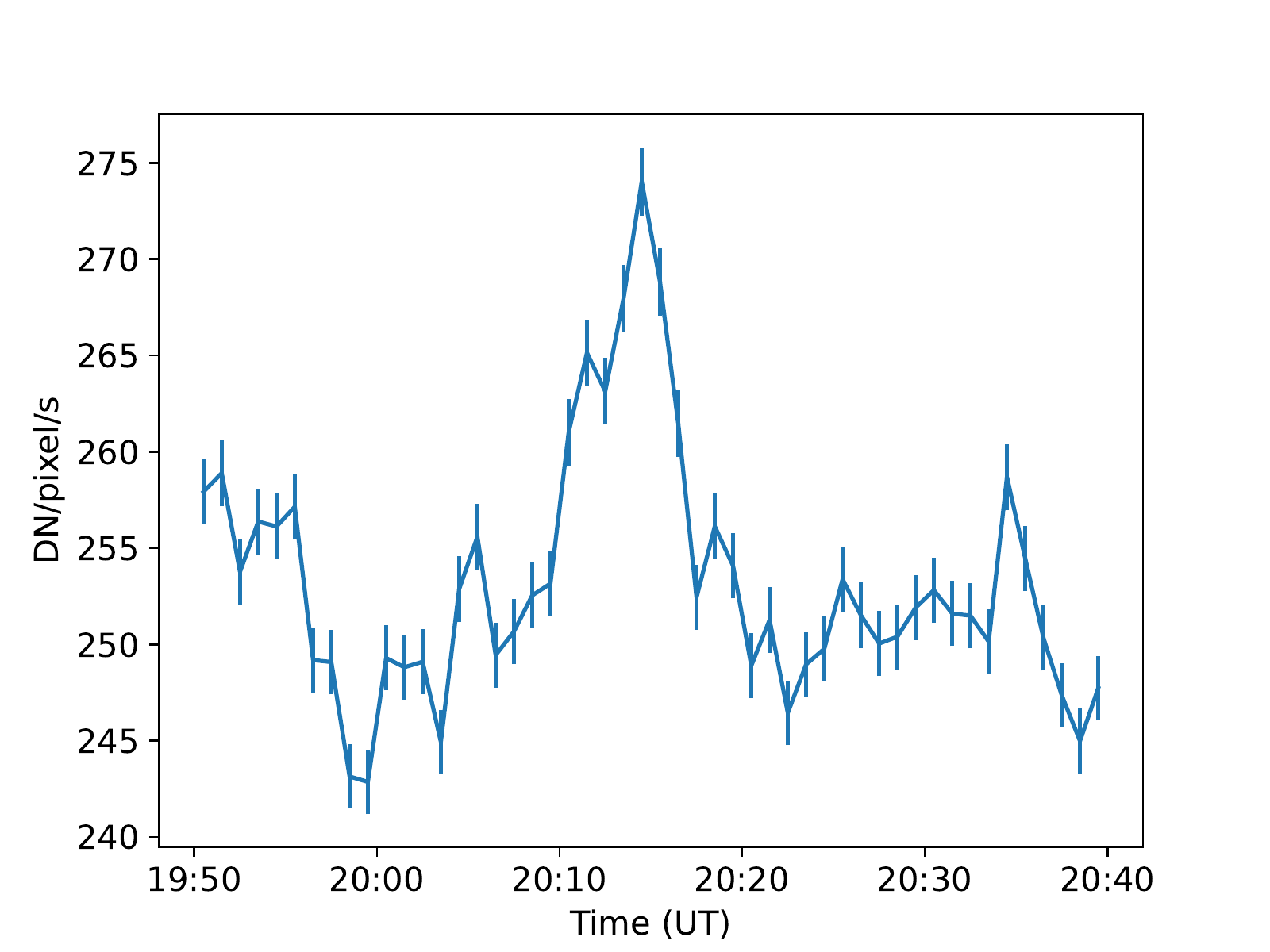}
       \includegraphics[trim={1.2cm 0 0 0},clip,scale=0.38]{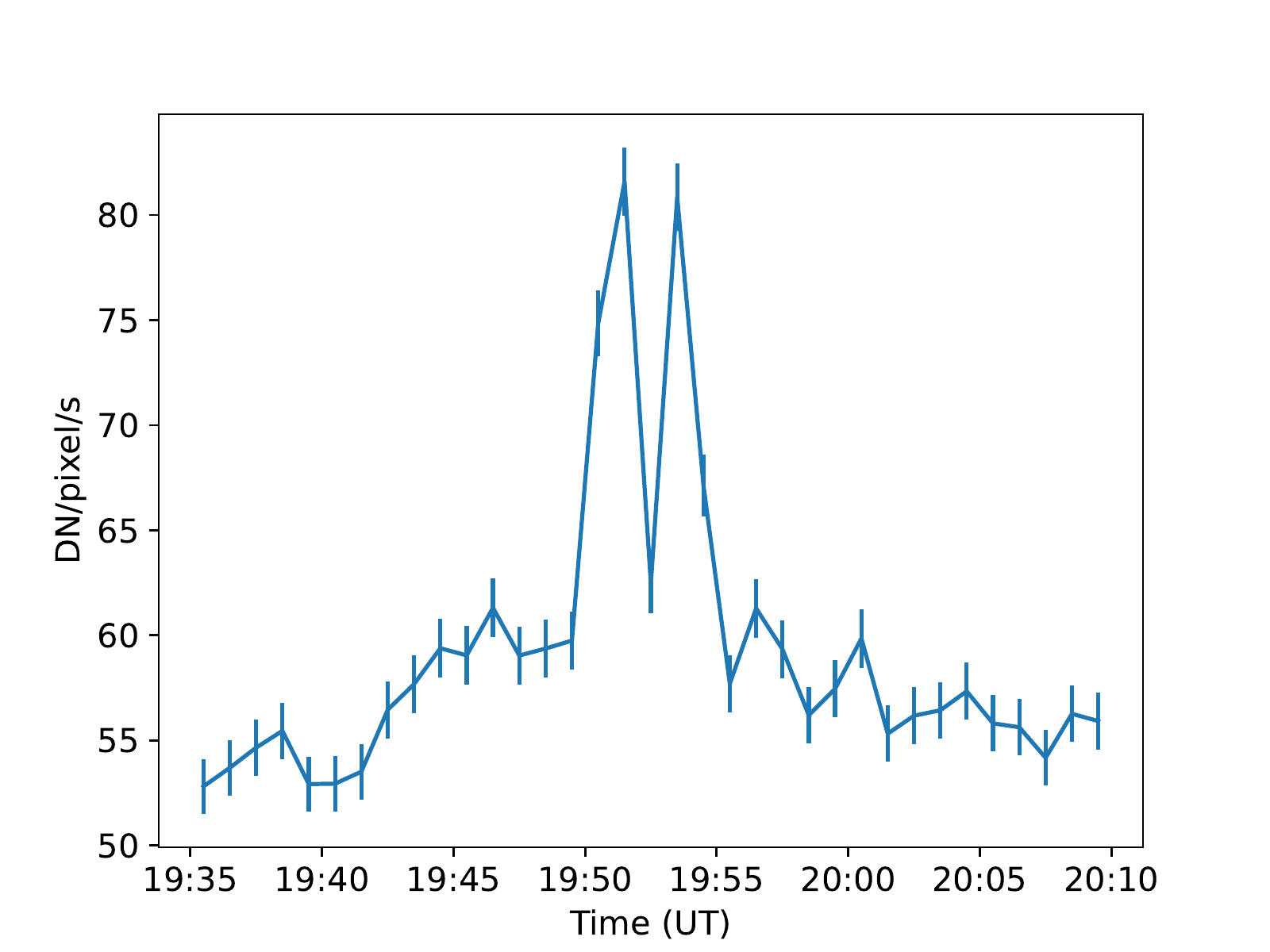}
       \includegraphics[trim={1.2cm 0 0 0},clip,scale=0.38]{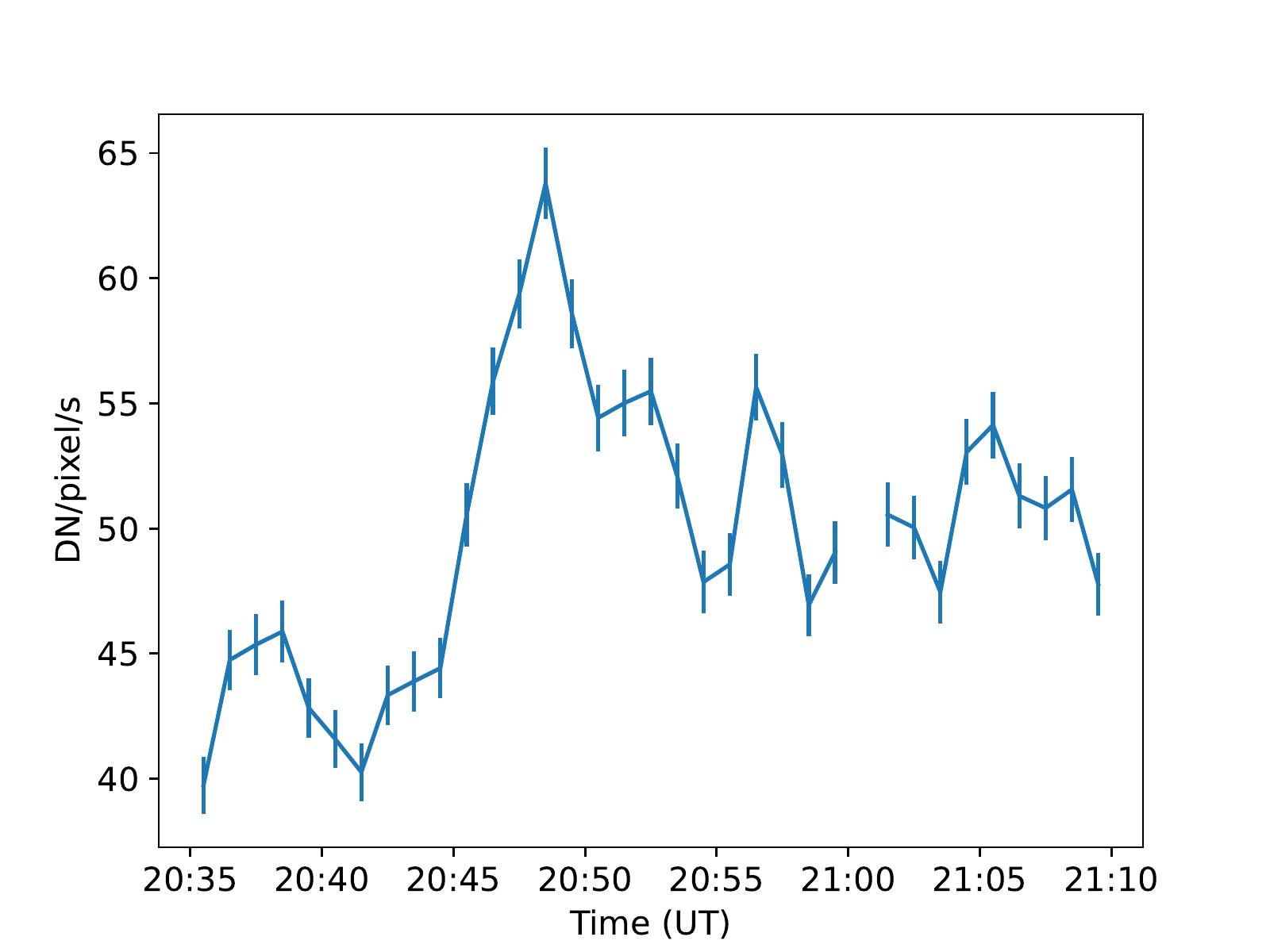}
       \includegraphics[scale=0.38]{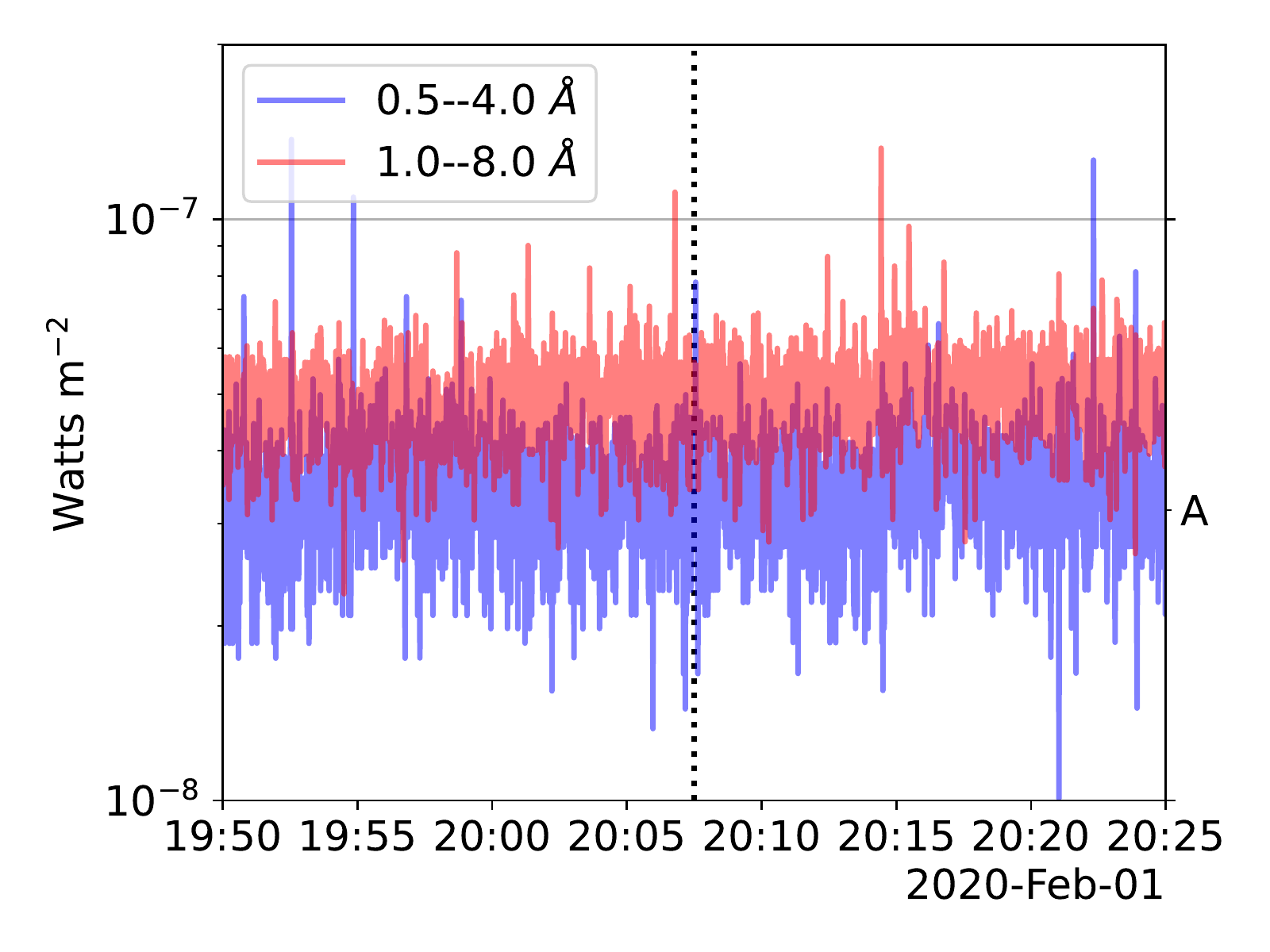}
       \includegraphics[trim={1.2cm 0 0 0},clip,scale=0.38]{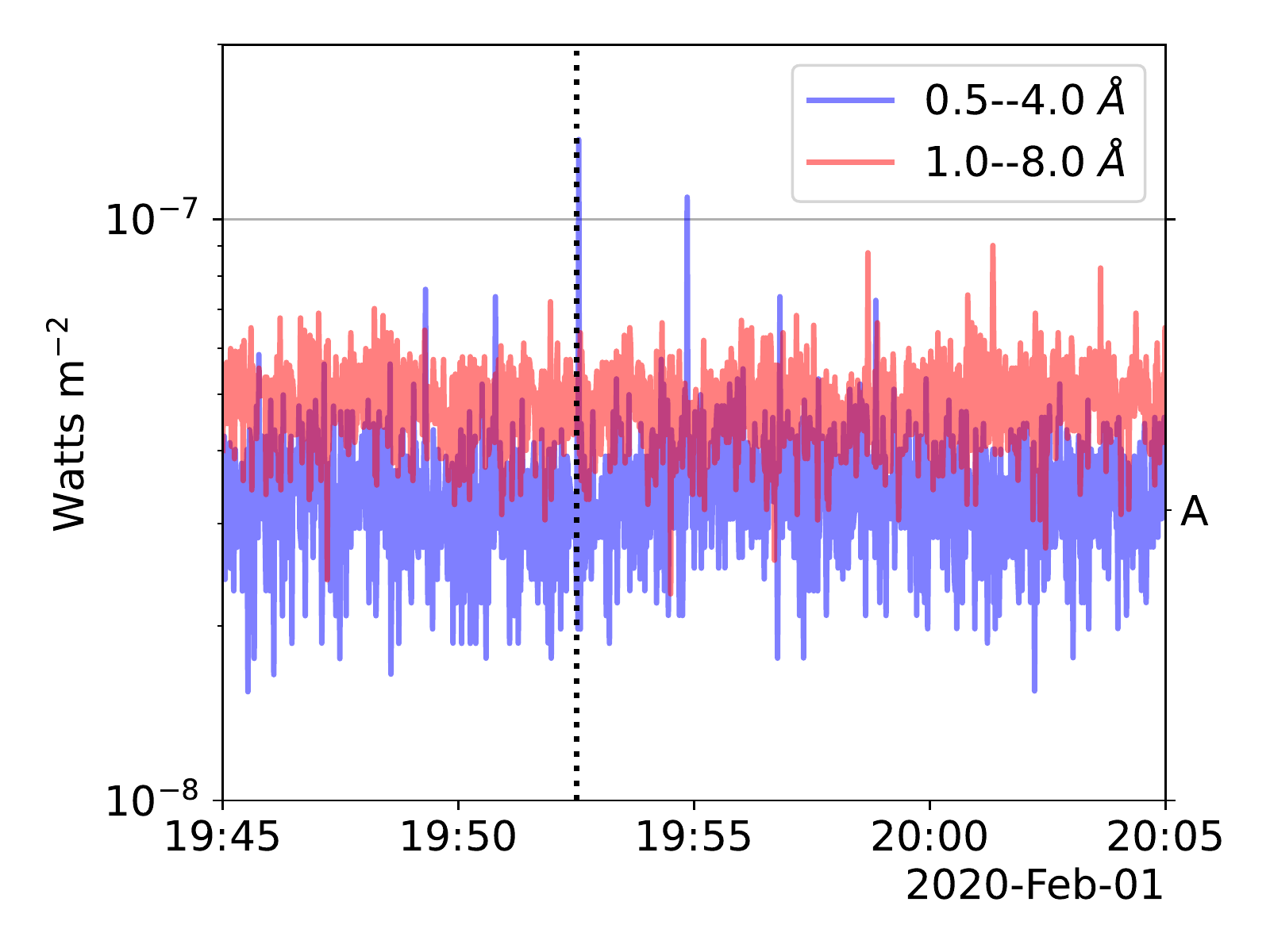}
       \includegraphics[trim={1.2cm 0 0 0},clip,scale=0.38]{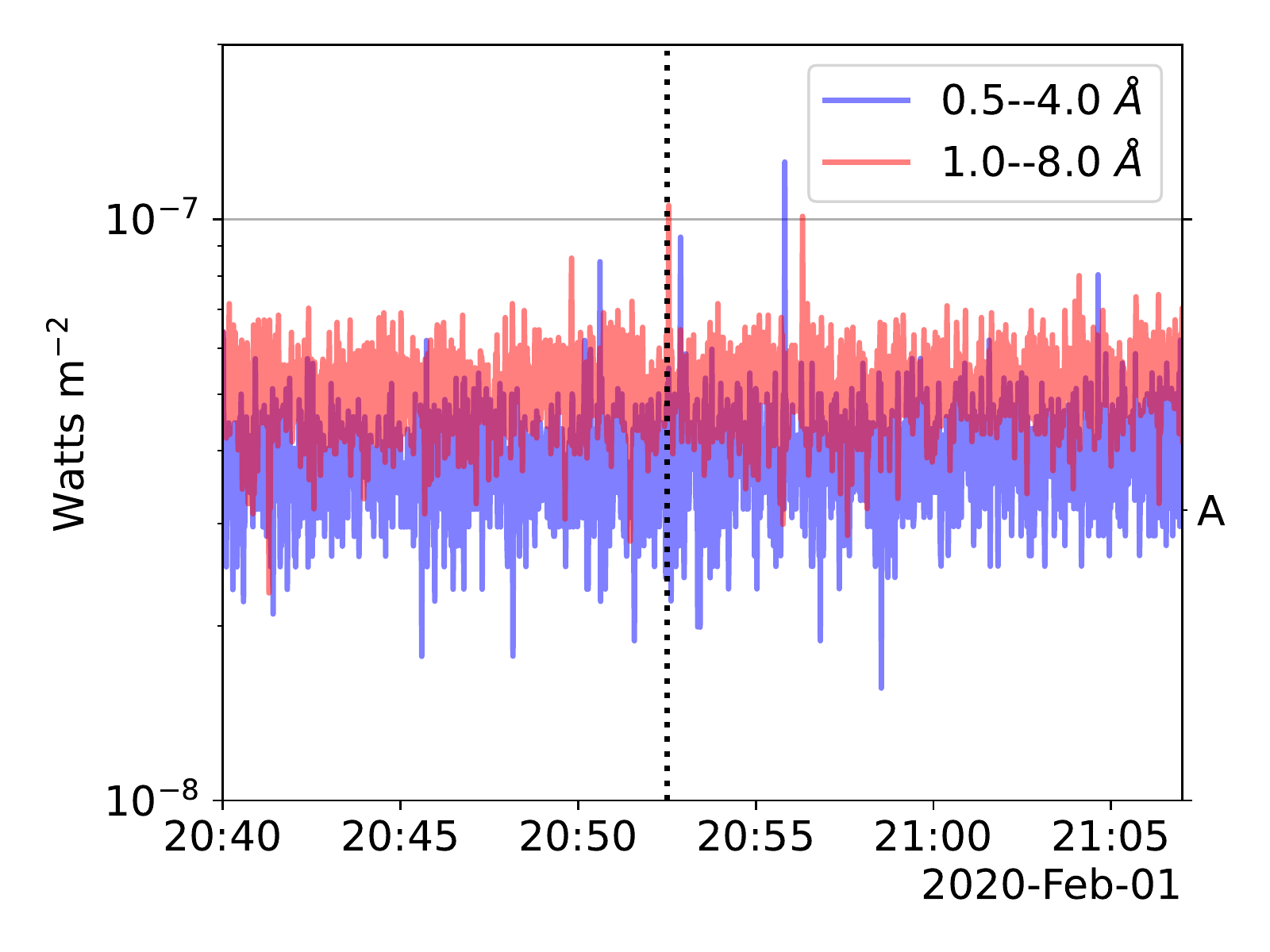}
    \caption{Radio transients in the coronal hole region detected by VLA. First row: Overlay of radio contours over SDO/AIA images at similar times. Second row: Light curves of radio sources at {1.313 (left panel) and 1.987 GHz (middle and right panels)}. The red points show the peak flux at the location of the radio source. The blue triangles show the 3$\sigma$ value, where $\sigma$ is the estimated noise on the corresponding radio image. Third row: Light curves of the box centered on the red dashed point shown in the upper panel in the SDO/171 \AA$\,$ image. {Fourth row: GOES lightcurve during the relevant times are shown. The dotted black line is used to mark the peak of the radio light curve.} Each column corresponds to the same source. }
    \label{fig:coronal_hole_sources}
\end{figure*}

\begin{figure}
    \begin{interactive}{animation}{fig007.mp4}
    \includegraphics[scale=0.5]{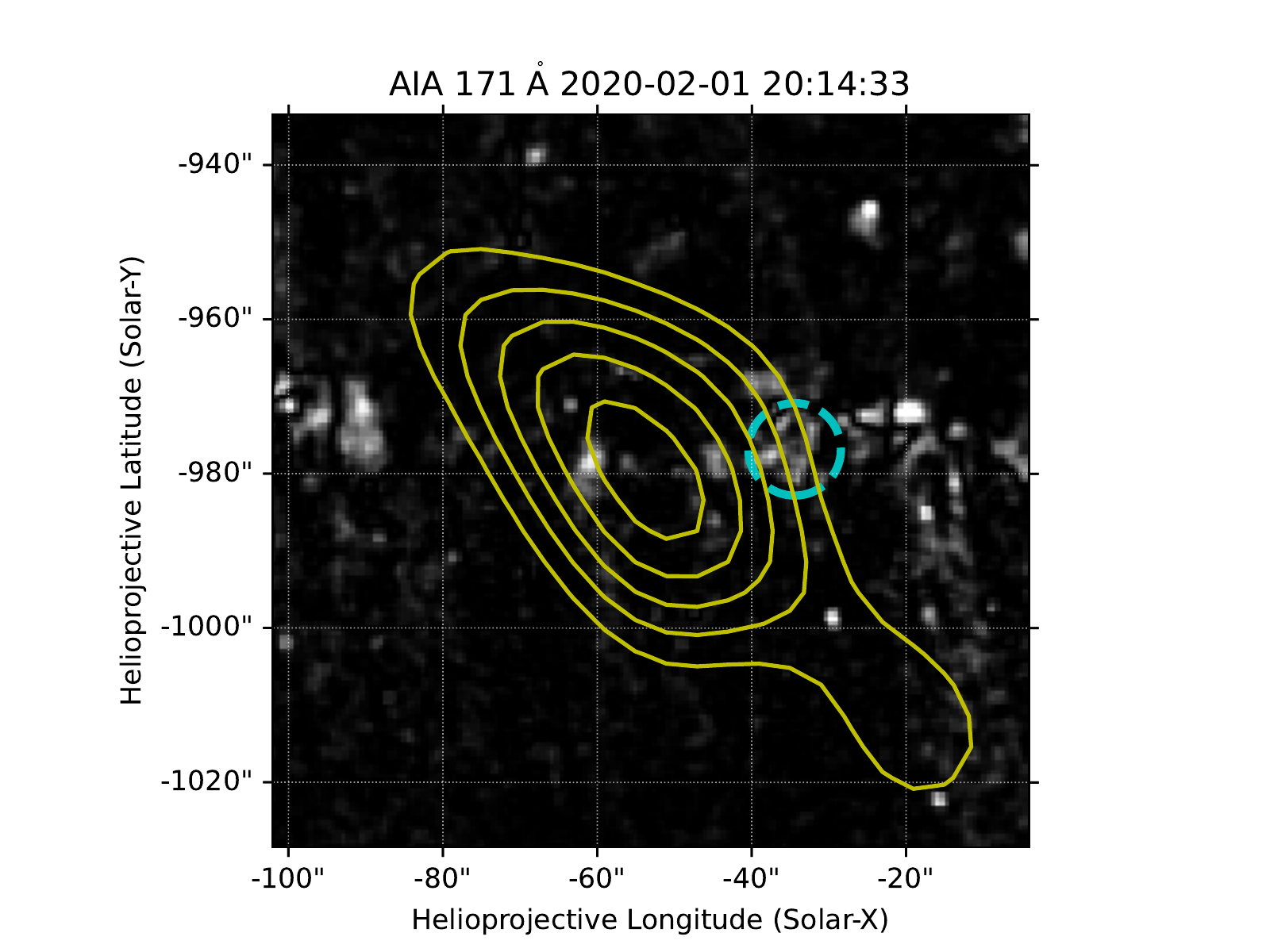}
    \end{interactive}
    \caption{Base difference EUV image at 171\AA$\,$  corresponding to the source shown the left panel of Fig. \ref{fig:coronal_hole_sources}. {This figure is available as an animation in the supplementary material. The animation shows the variability seen in the difference image with time. The animation spans from 19:54:00--20:27:00 UT and shows the AIA images at a cadence of 1 minute.}}
    \label{fig:spw2_coronal_hole_source1}
\end{figure}

\begin{figure}
    \centering
    \begin{interactive}{animation}{fig008.mp4}
    \includegraphics[scale=0.5]{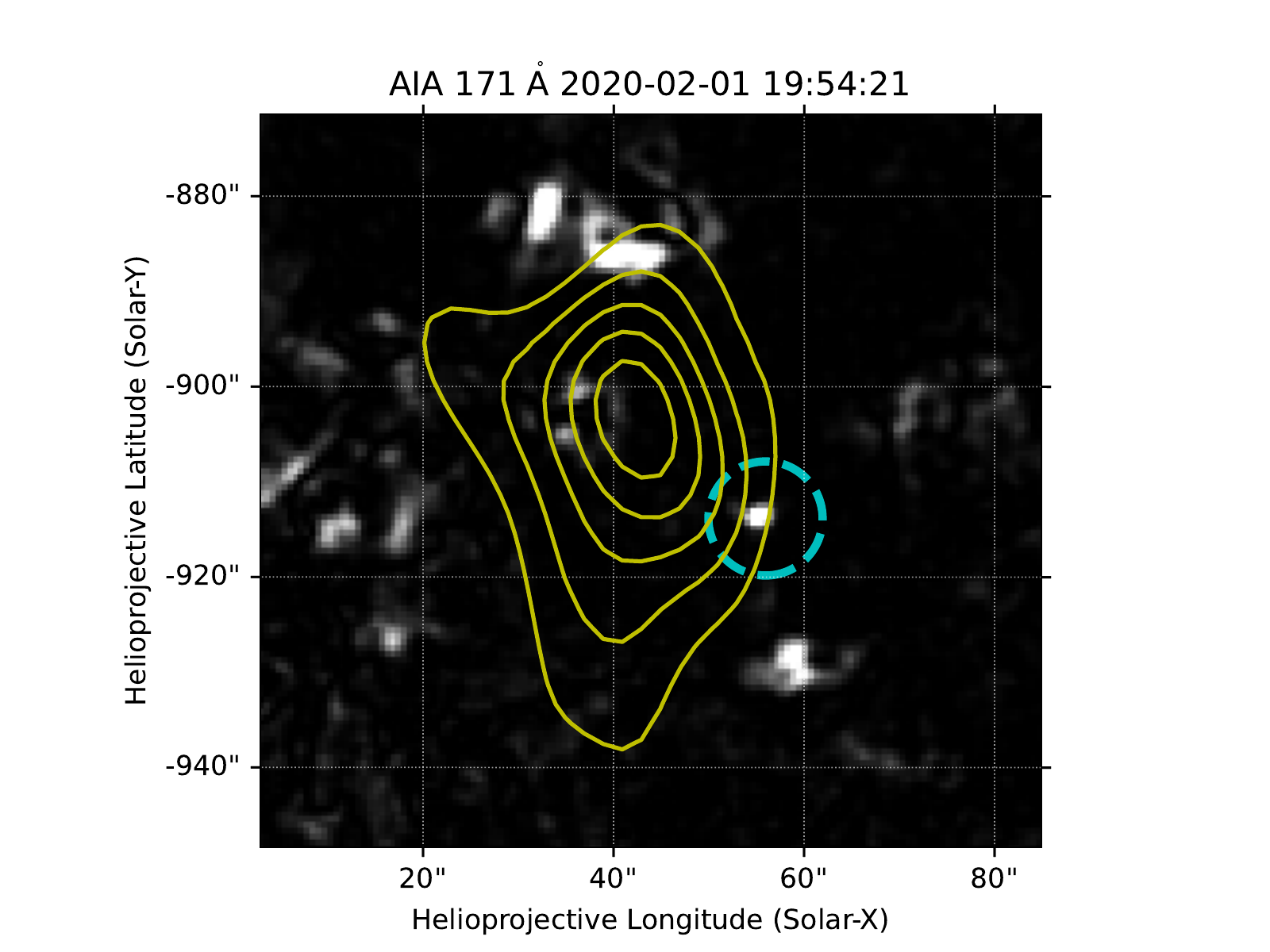}
    \end{interactive}
    \caption{Base difference EUV image at 171\AA$\,$ corresponding to the source shown in the middle panel of Fig. \ref{fig:coronal_hole_sources}. {This figure is available as an animation in the supplementary material. The animation shows the variability seen in the difference image with time. The animation spans from 19:35:00--20:10:00 UT and shows the AIA images at a cadence of 1 minute.}}
    \label{fig:spw7_coronal_hole_source1}
\end{figure}

\begin{figure}
    \centering
    \begin{interactive}{animation}{fig009.mp4}
    \includegraphics[scale=0.5]{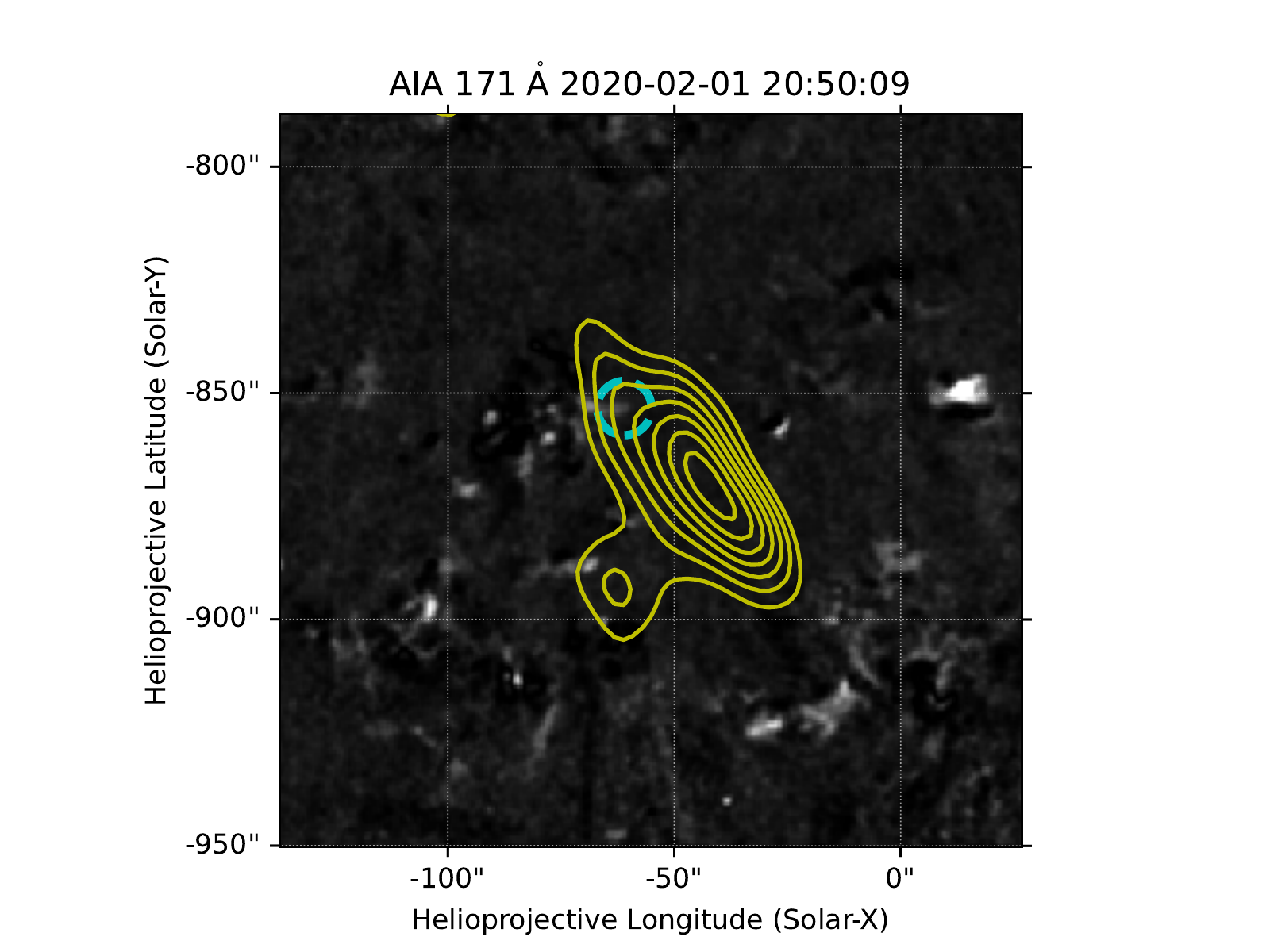}
    \end{interactive}
    \caption{Base difference EUV image at 171\AA$\,$ corresponding to the source shown in the right panel of Fig. \ref{fig:coronal_hole_sources}. {This figure is available as an animation in the supplementary material. The animation shows the variability seen in the difference image with time. The animation spans from 20:35:00--21:10:00 UT and shows the AIA images at a cadence of 1 minute.}}
    \label{fig:spw7_coronal_hole_source2}
\end{figure}

While several sources were detected in coronal holes, three sources were chosen for further analysis because these are the brightest sources in the respective images and hence were unlikely to be significantly affected by the sidelobes of other sources. The radio contours of these three sources have been overlaid on SDO/AIA images at a similar time, and are shown in the first row of Fig. \ref{fig:coronal_hole_sources}. {The sources in the right, middle, and left panels are detected at 1.313, 1.941, and 1.941 GHz respectively.} In the second and third rows, the light curves of these radio sources and those of the EUV sources (centers of the cyan dashed circles drawn in the upper panels) are shown. The EUV sources have been identified through visual inspection such that they are located close to the radio source and their light curve shows qualitative similarity with that of the corresponding radio source. However, it is quite possible that multiple EUV sources lying within the instrumental resolution element of the radio instruments may contribute to the observed radio emission. Hence while identification of the EUV source bolsters the fidelity of the detected sources, their exact location is only accurate up to the instrumental resolution of the radio data. In the fourth row, the GOES light curve is shown. The black dotted line shows the time when the radio flux density is maximum. It should be noted that while this line has a spread of 5 minutes, which is the integration time of the radio images, the spread has not been shown in these plots. Interestingly we find that for each source, we find an X-ray peak at the same location as the radio peak. However, it should also be noted that at these low flux levels GOES light curve has a lot of contamination due to electrons and in absence of independent confirmation, the mere presence of these X-ray peaks does not confirm an X-ray signature. {Movies showing the variability in EUV at the location of these three sources are provided in the supplementary material. , where we overlay contours of these three radio images over AIA 171\AA$\,$ base difference images. The cyan circles shown are the same ones shown in the upper panel of Fig. \ref{fig:coronal_bright_point_source}. The movies span the same time duration as the light curves shown in the second row of Fig. \ref{fig:coronal_bright_point_source}, with a cadence of 1 minute. Still frames of these movies are given in Figs. \ref{fig:spw2_coronal_hole_source1}, \ref{fig:spw7_coronal_hole_source1} and \ref{fig:spw7_coronal_hole_source2}. The name of the movie is same as the respective figure number of its still frame.}

\begin{figure}
    \centering
    \includegraphics[scale=0.5]{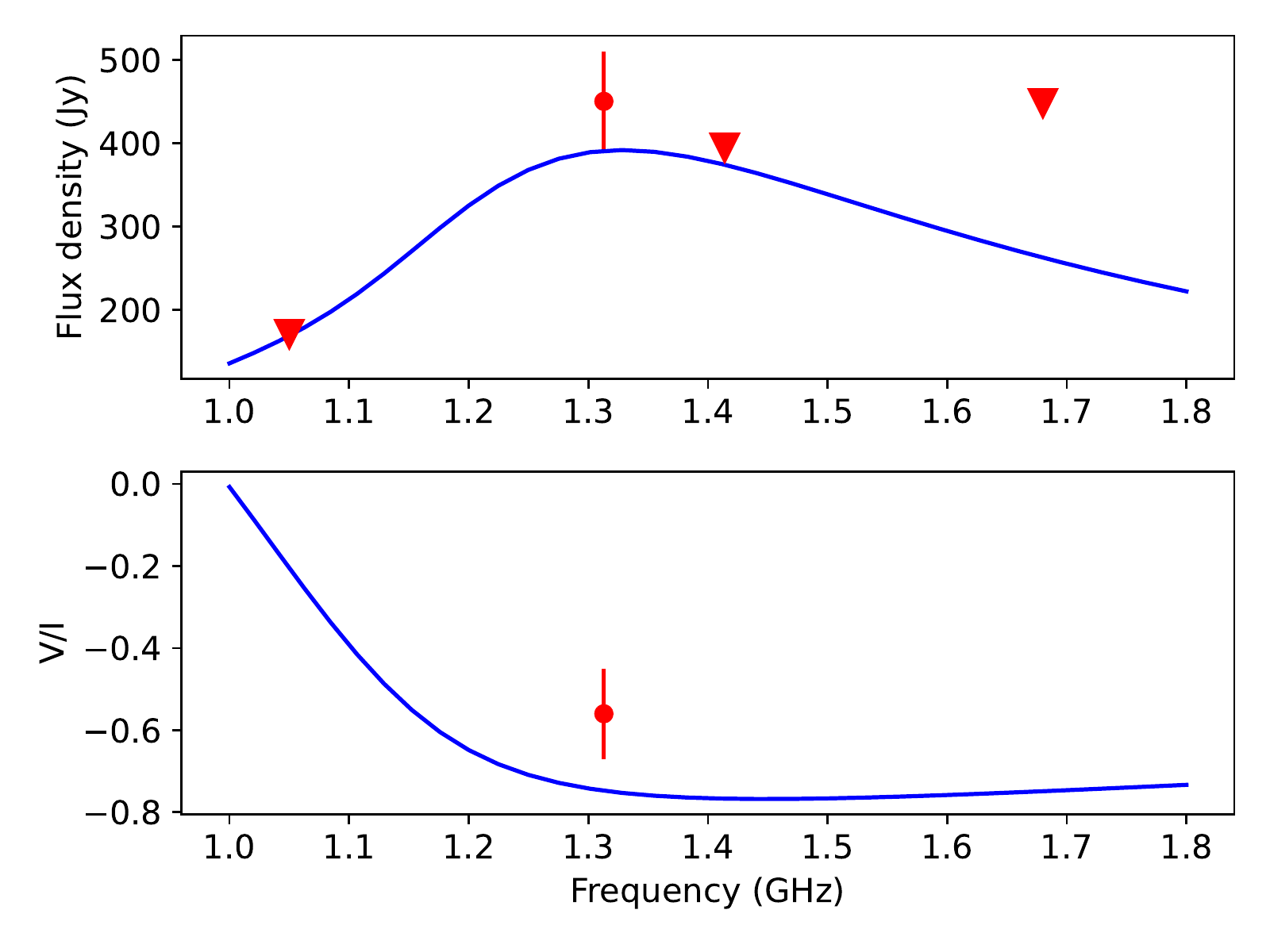}
    \caption{Flux density spectrum of the source in the left panel of Fig. \ref{fig:coronal_hole_sources}. The red circular points shows the observed values. The red triangles show the upper limits to the flux density estimated as 5 times the noise in the image. The blue line shows a representative gyrosynchrotron model which satisfies the observational constraints. }
    \label{fig:coronal_hole_source1_qualitative}
\end{figure}

We find that the radio sources are located near the coronal bright points in the coronal hole region. From the top panel, it is evident that the EUV source is co-located with the radio source and also shows variability in similar times. The variability is always observed in the 171\AA$\,$, and sometimes in the 193\AA$\,$, and 211\AA$\,$, but not observed in the other AIA wavebands.  { It should however be noted that due to the huge difference between the instrumental resolution of AIA and the radio observations presented here, it is extremely hard to identify a unique EUV counterpart to the radio source and it is possible that the observed radio emission has a contribution from other nearby sources as well.} 
All of these radio sources are circularly polarized, with the one in the left column showing a degree of circular polarization of $-56\pm 11\%$. The sources in the middle and right columns are only detected in the right circular polarization and using the 5$\sigma$ as the upper limit of the flux in the left circular polarization we estimate that the stokes V fraction is greater than 11\% and 23\% respectively. The high circular polarization of the source can be explained either by optically thin free-free emission due to propagation effects through the magnetic field \citep[e.g.][]{habbal1986,sastry2009}, or due to gyrosynchrotron emission. {Extremely impulsive narrowband emissions with flux densities comparable to the sources studied here were detected only in one circular polarisation \citep{bastian1991} and were attributed to plasma emission mechanism. While those emissions were solely associated with active regions, plasma emission from the weak reconnection events happening in the quiet Sun has been reported at lower frequencies \citep{mondal2020,mondal2023,sharma2022} and can happen at our observing frequencies as well.} Hence the polarization data alone are insufficient to determine the emission mechanism of these sources. However, for the first source shown in the left column, we can rule out free-free emission by combining the polarization and spectral information. Fig. \ref{fig:coronal_hole_source1_qualitative} shows the observed flux density and the corresponding degree of circular polarization at 1.313 GHz where we have a detection of confidence. The upper limits of the flux density of the source at neighboring frequencies are also computed as 5 times the noise in the image, shown as red triangles. While the source spectra (observed flux at 1.313 GHz and the upper limits at other frequencies) may be explained by optically thick free-free emission, the high degree of circular polarization cannot be explained. On the other hand, while the polarization signal can be explained with optically thin free-free emission due to propagation effects, the spectral structure cannot be explained as the spectrum is not flat. However, a gyrosynchrotron model can explain both properties simultaneously. A representative gyrosynchrotron spectrum which satisfies the spectral constraints and also shows a high degree of circular polarization is shown as the blue curve in Fig. \ref{fig:coronal_hole_source1_qualitative}. { Note that this model spectrum shown here is not a best-fit spectrum due to the lack of spectral constraints. It is provided with the sole intention of demonstrating that a nonthermal gyrosynchrotron spectrum can possibly meet the observational constraints in a much better manner than a thermal free-free emission mechanism.}

\subsection{Sources associated with network regions} \label{sec:network}

\begin{figure*}
    \centering
    \includegraphics[scale=0.48]{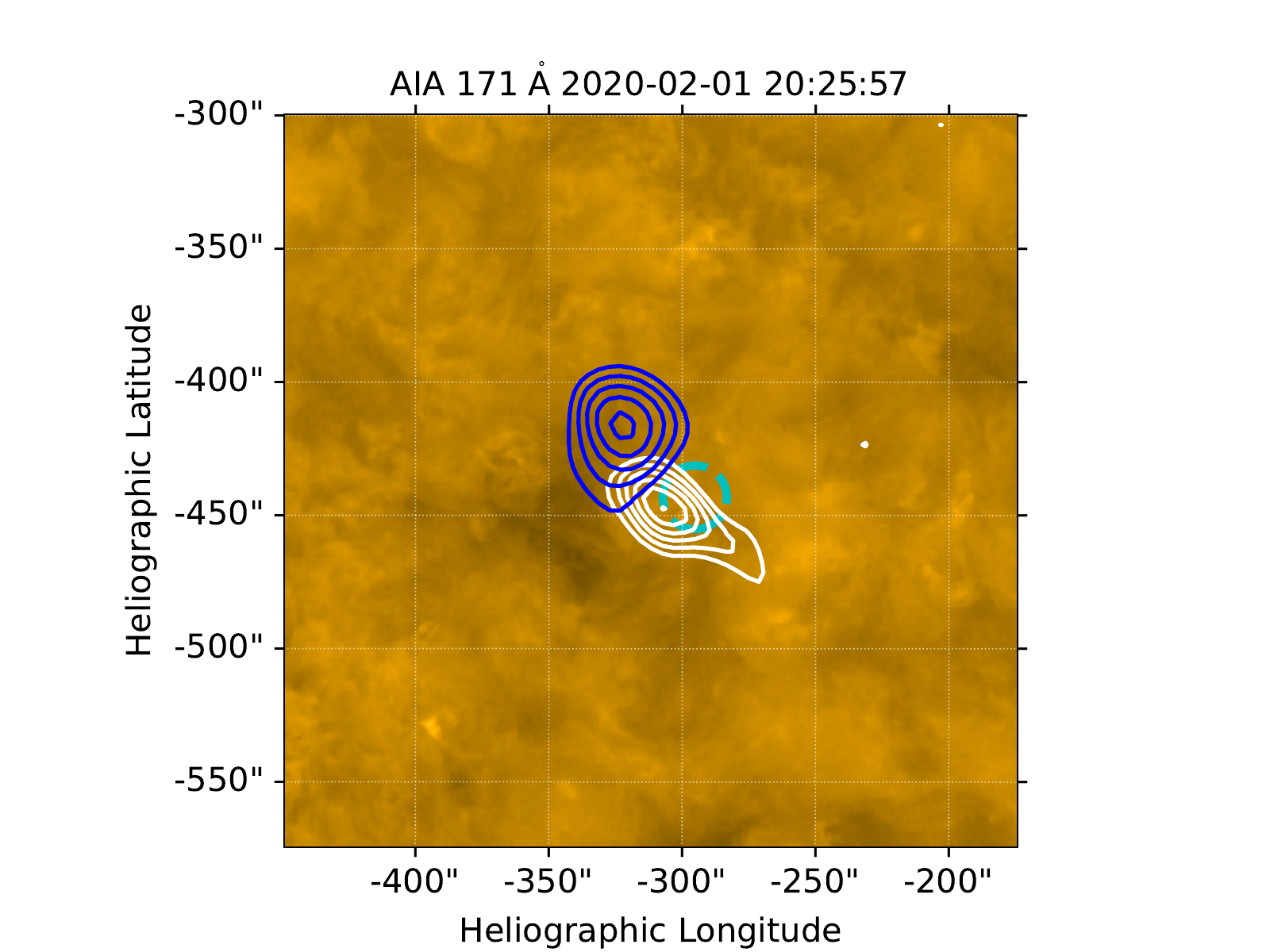}
    \includegraphics[trim={2cm 0 0 0},clip,scale=0.48]{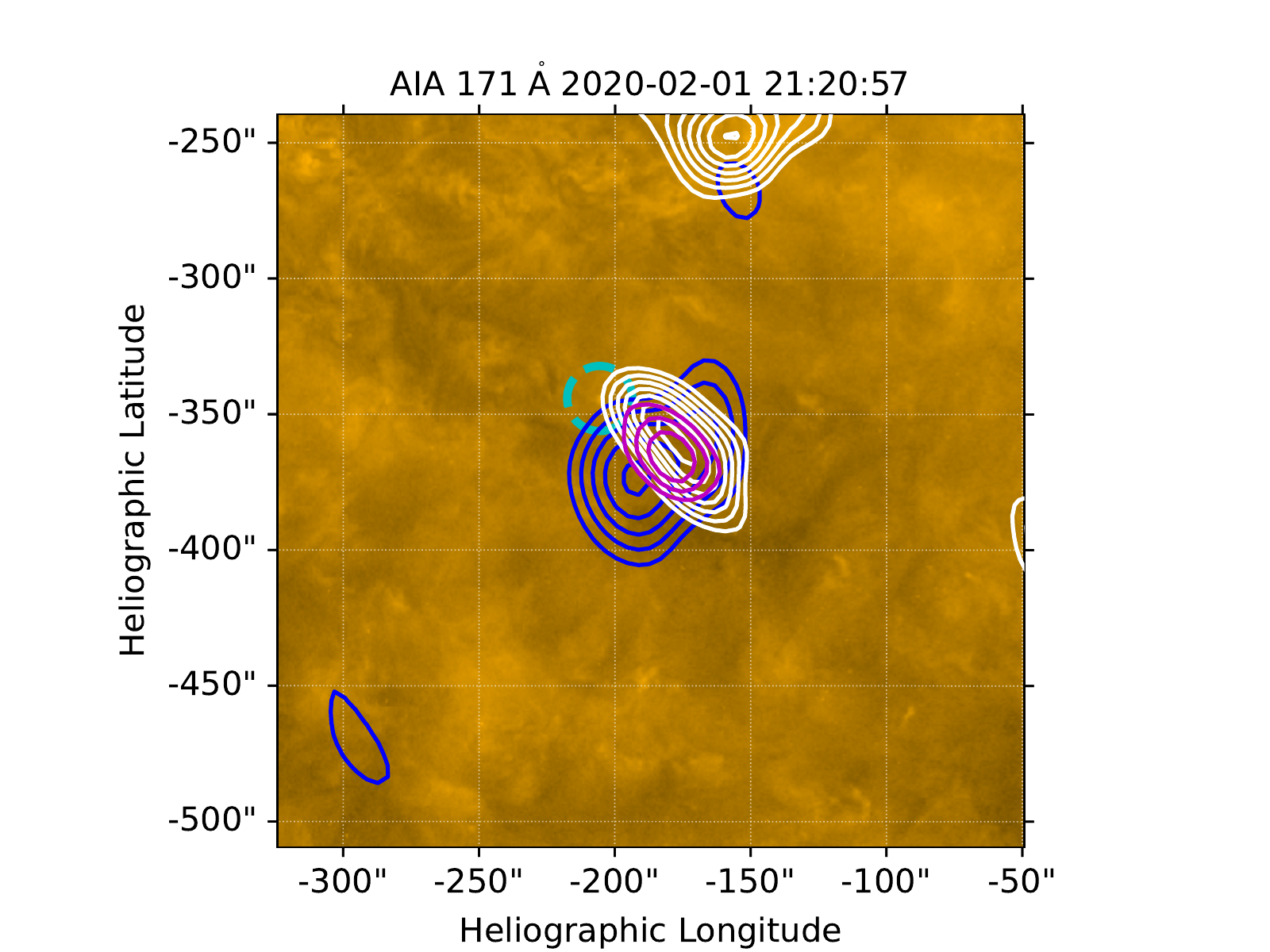}
    \includegraphics[scale=0.4]{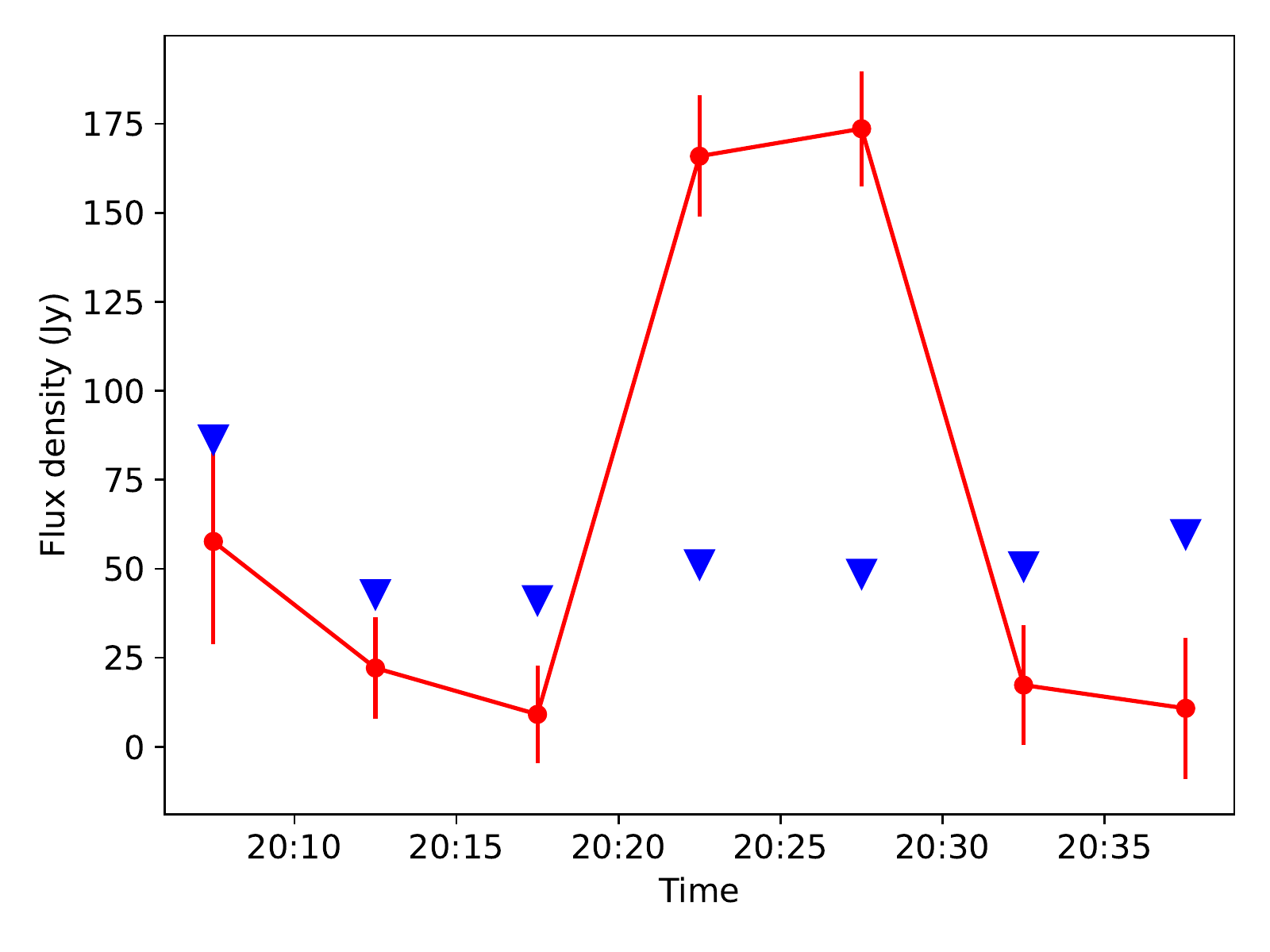}
    \includegraphics[scale=0.4]{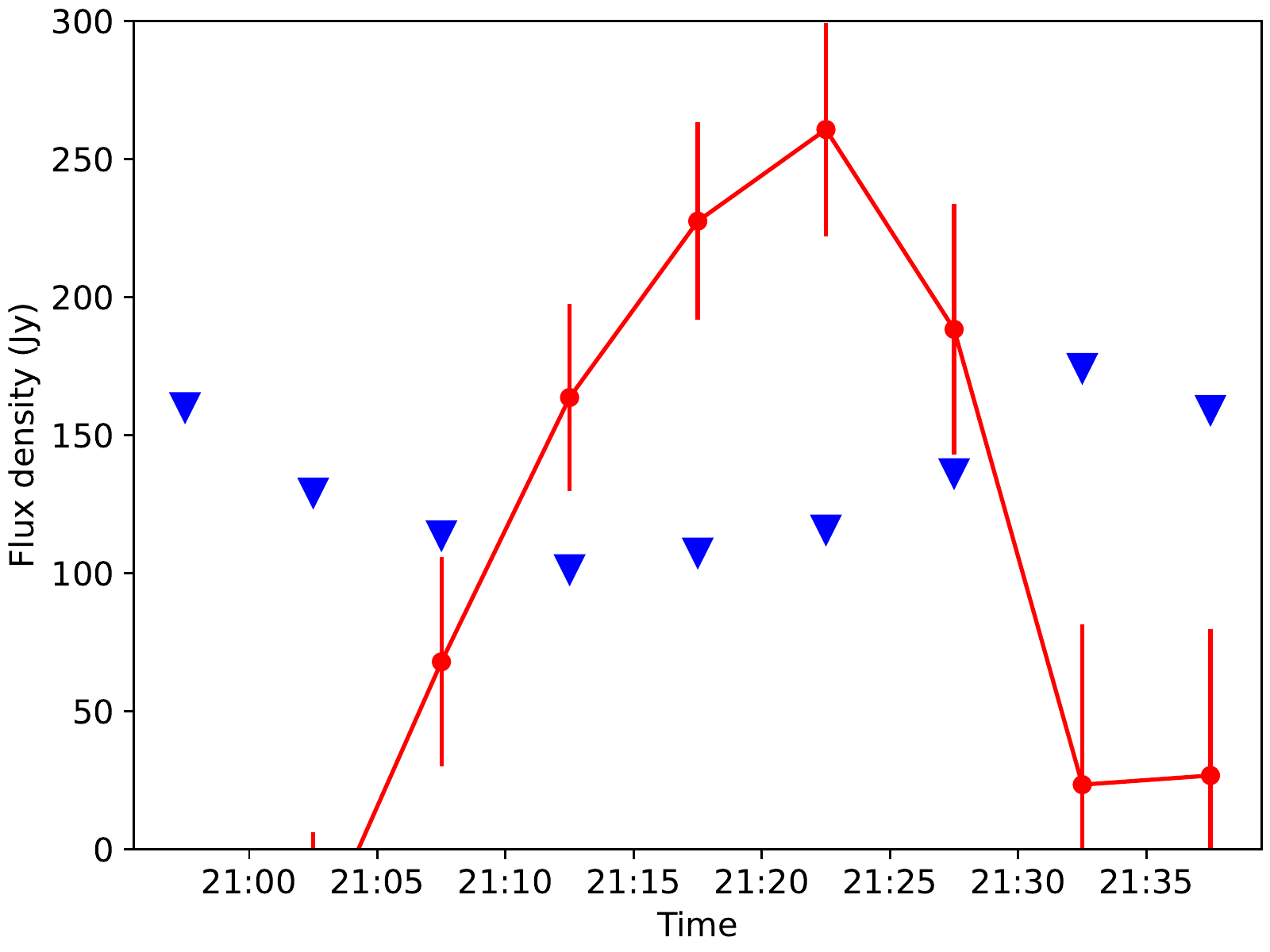}
     \includegraphics[scale=0.4]{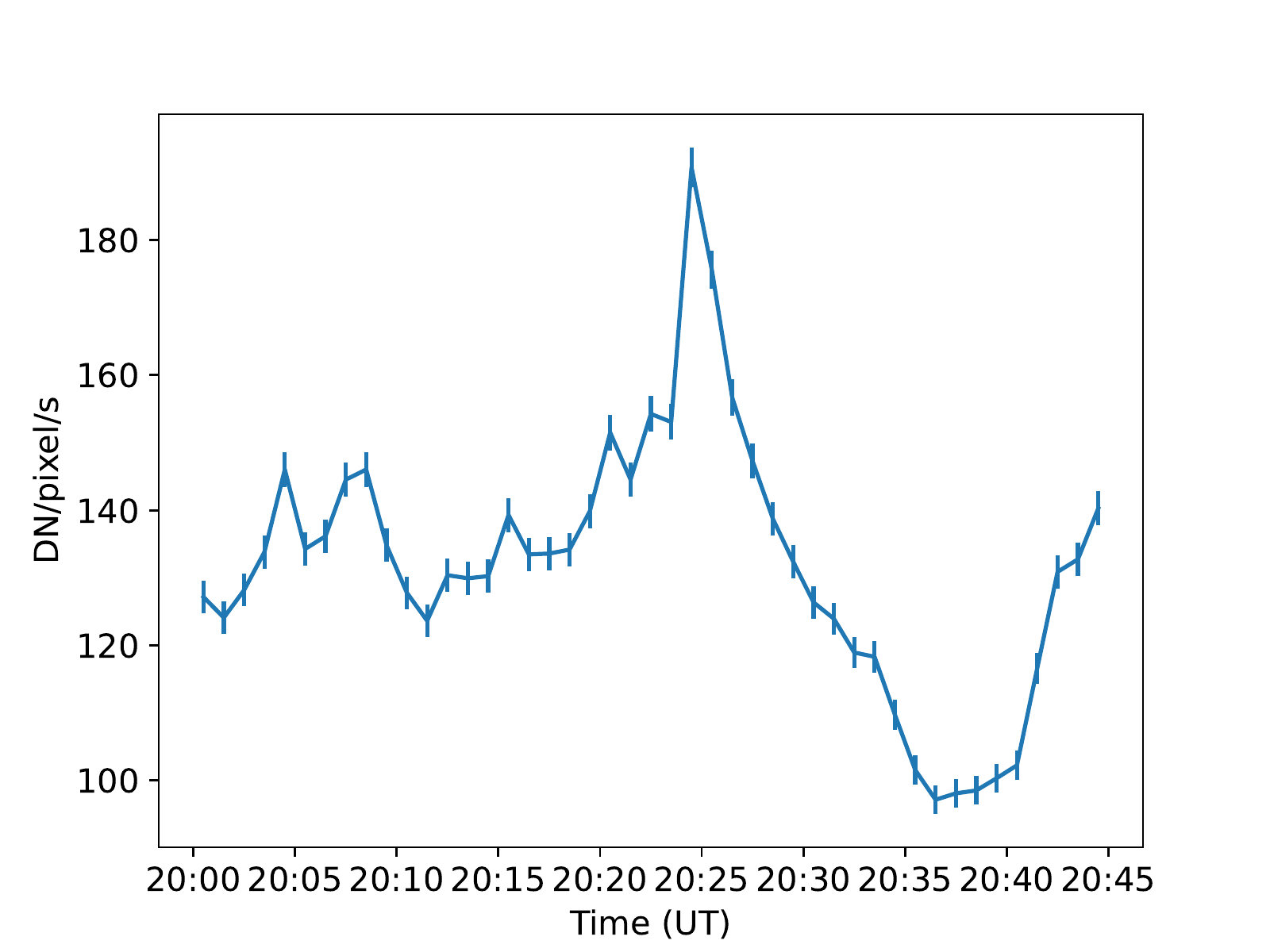}
     \includegraphics[trim={0.9cm 0 0 0},clip,scale=0.4]{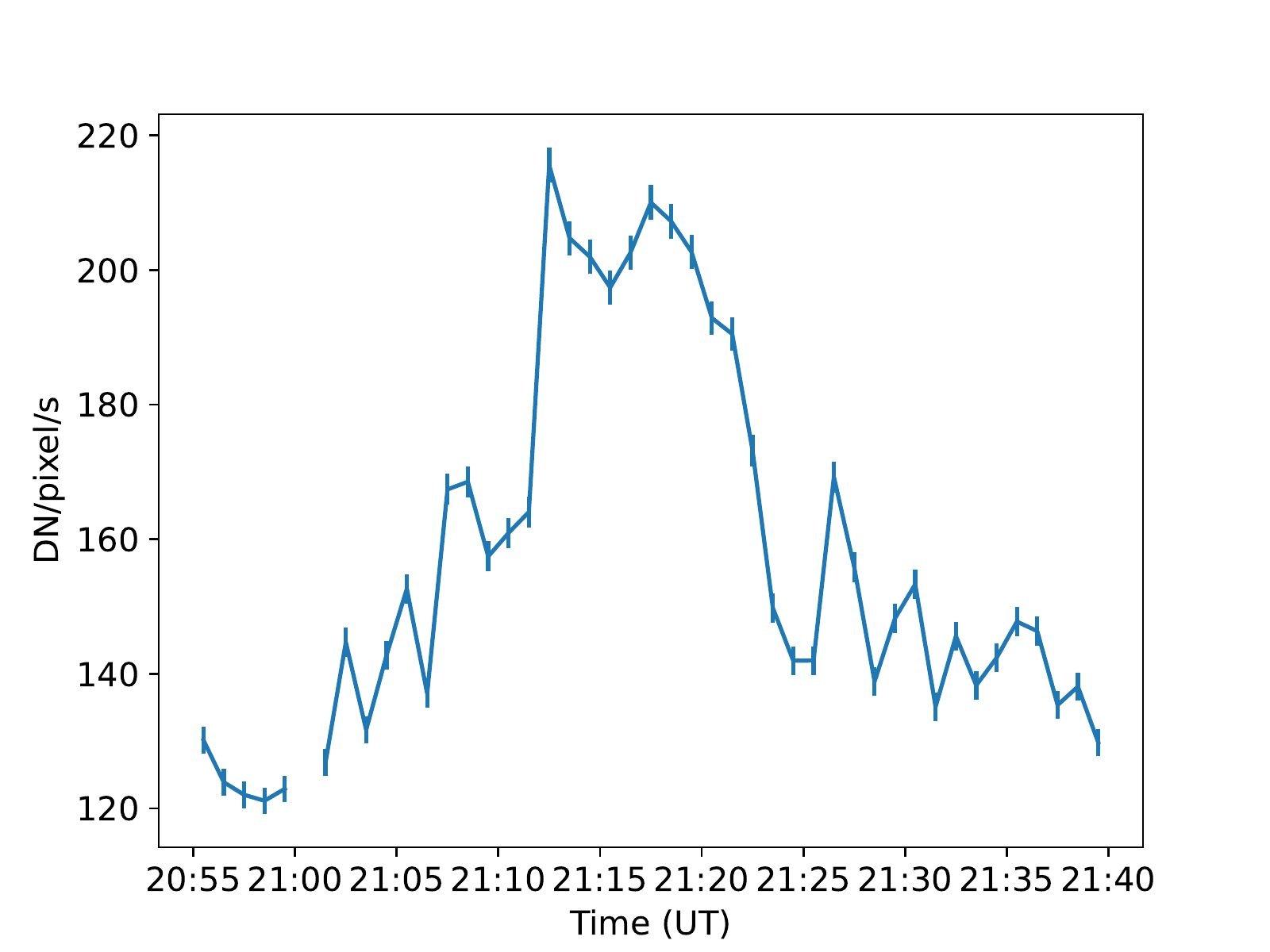}
     \includegraphics[scale=0.4]{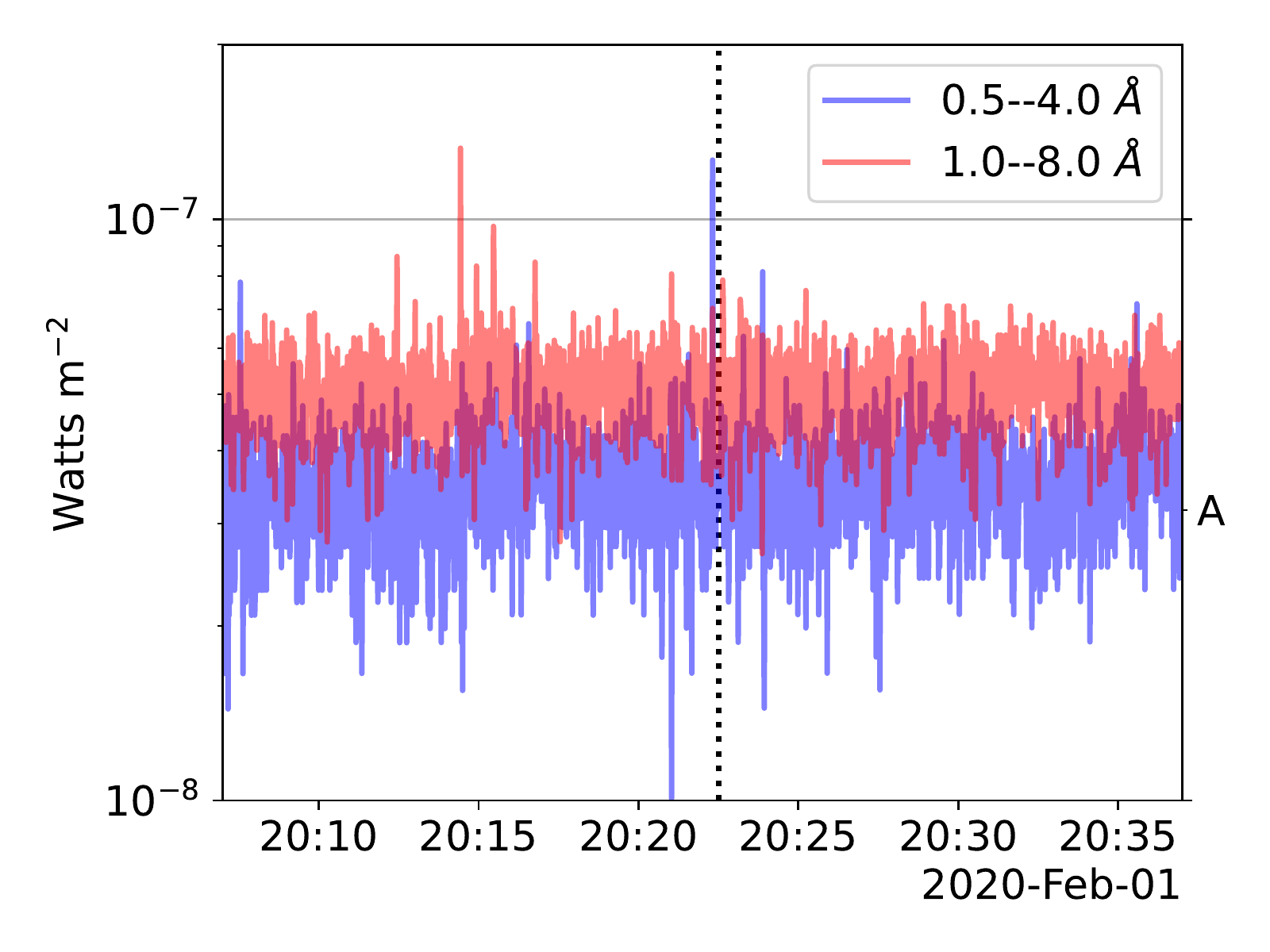}
     \includegraphics[trim={0.9cm 0 0 0},clip,scale=0.4]{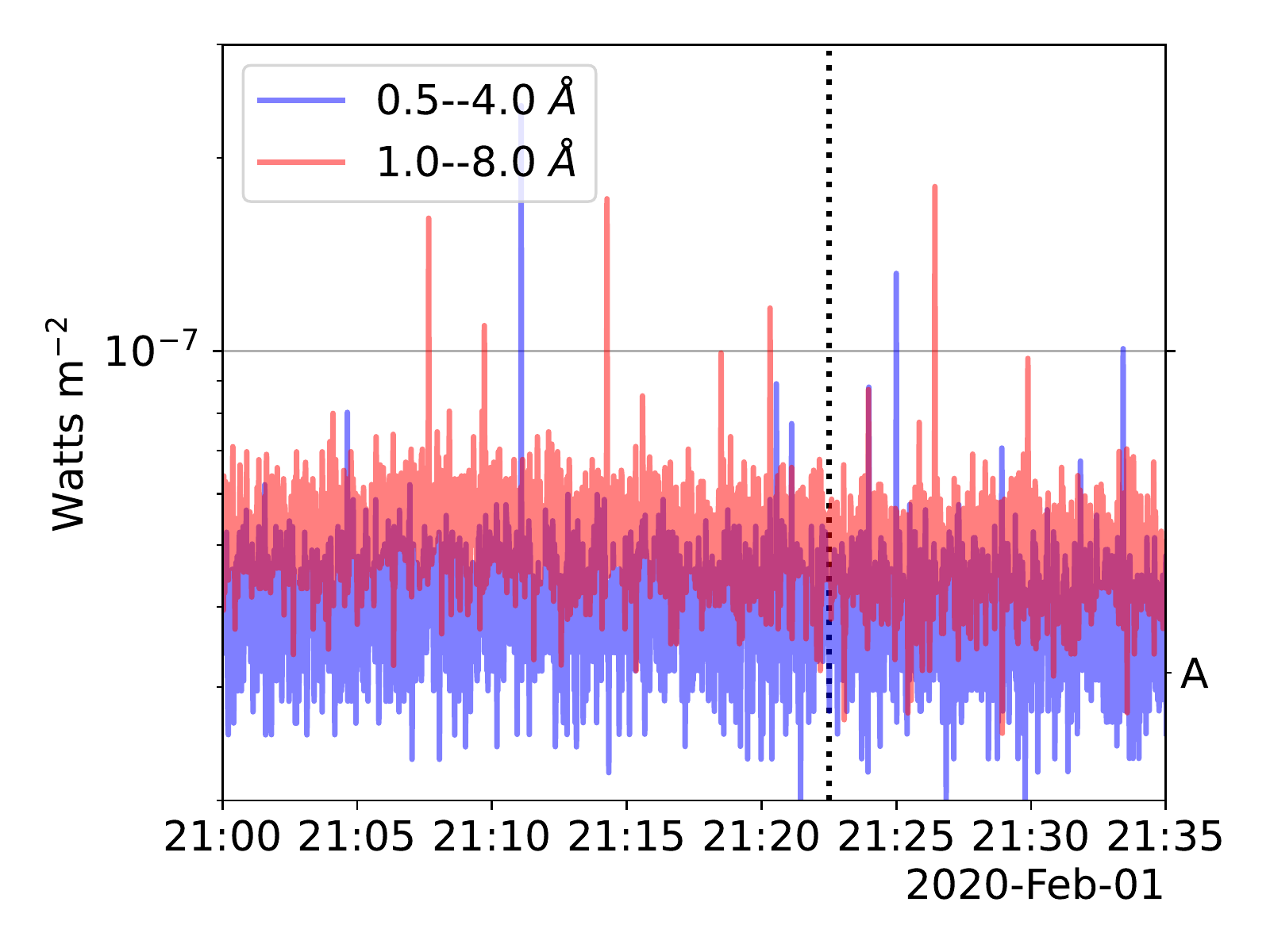}
    \caption{Image shown is in same format as that in Fig. \ref{fig:coronal_hole_sources} for the sources described in Section \ref{sec:network}. The left panel correspond to the source detected both in VLA and EOVSA images and the right panel correspond to the source detected in multiple frequencies.}
    \label{fig:network_sources}
\end{figure*}

\begin{figure}
  \begin{interactive}{animation}{fig012.mp4}
    \includegraphics[scale=0.5]{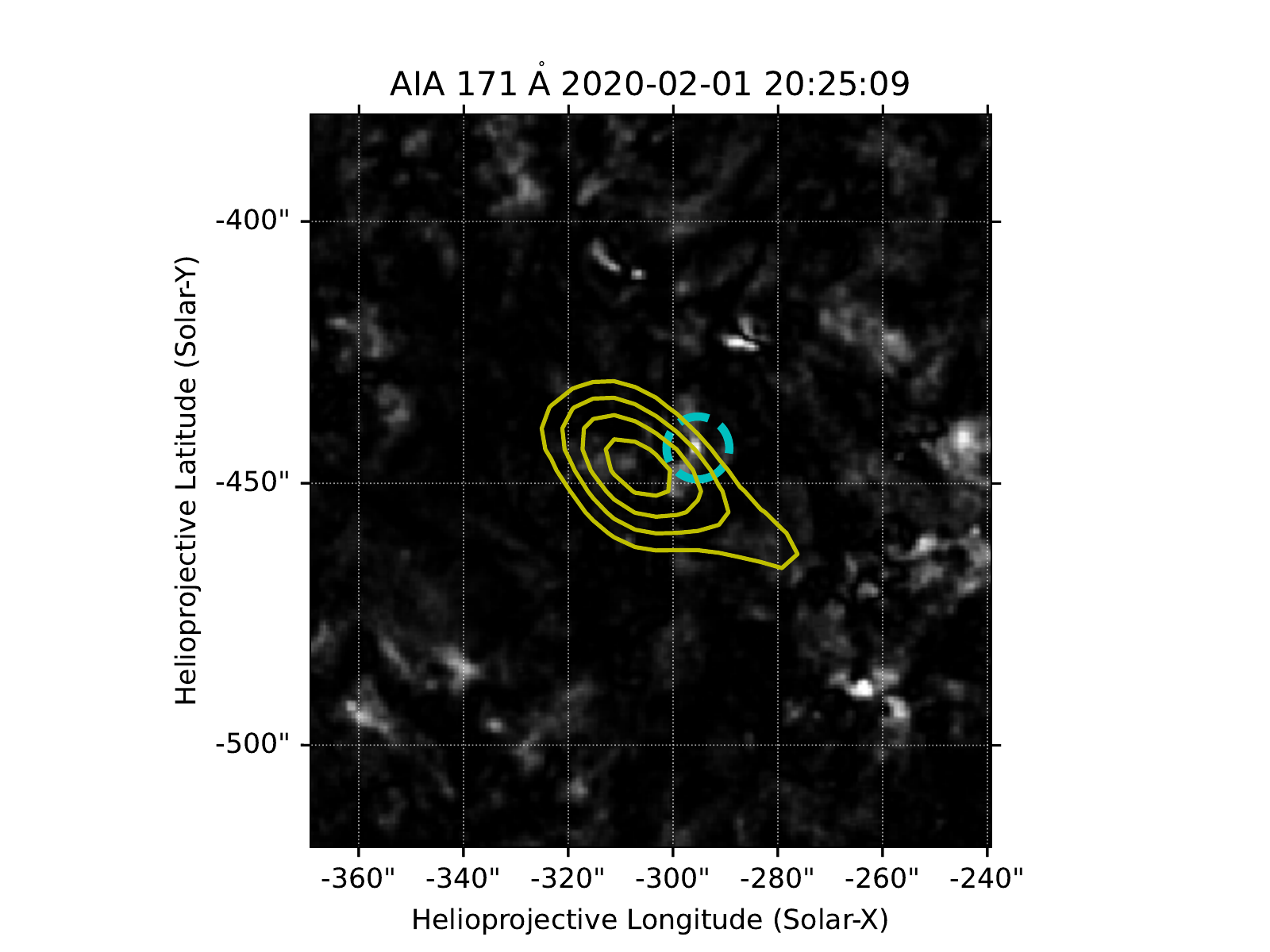}
    \end{interactive}
    \caption{Base difference EUV image at 171\AA$\,$ corresponding to the source shown in the left panel of Fig. \ref{fig:network_sources}. {This figure is available as an animation in the supplementary material. The animation shows the variability seen in the difference image with time. The animation spans from 20:00:00--20:45:00 UT and shows the AIA images at a cadence of 1 minute.}}
    \label{fig:spw5_network1}
\end{figure}

\begin{figure}
    \centering
    \begin{interactive}{animation}{fig013.mp4}
    \includegraphics[scale=0.5]{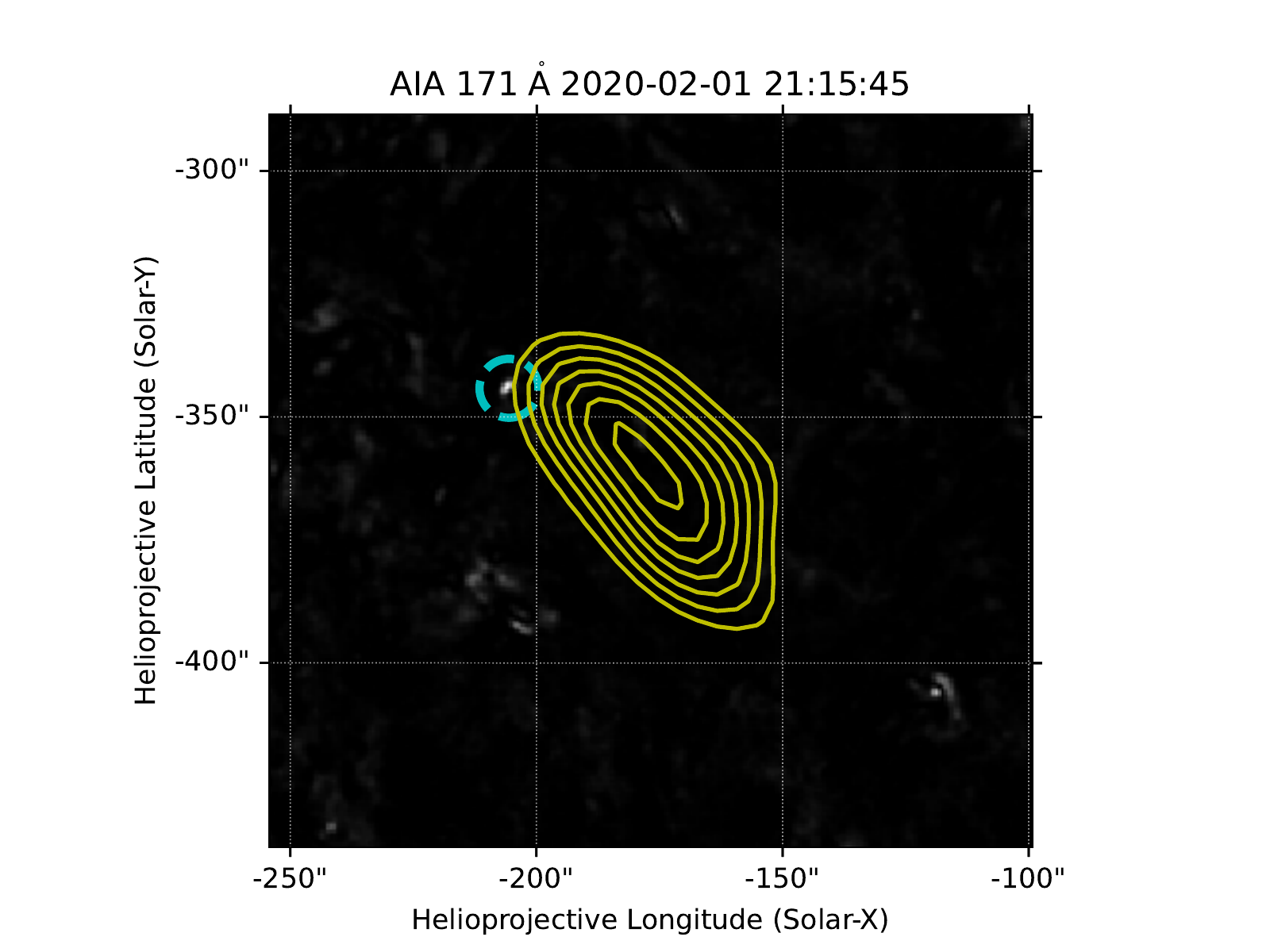}
    \end{interactive}
    \caption{Base difference EUV image at 171\AA$\,$ corresponding to the source shown in the right panel of Fig. \ref{fig:network_sources}. {This figure is available as an animation in the supplementary material. The animation shows the variability seen in the difference image with time. The animation spans from 20:55:00--21:40:00 UT and shows the AIA images at a cadence of 1 minute.}}
    \label{fig:spw2_network1}
\end{figure}

Several other radio transients were detected in the quiet sun network regions where no apparent dynamic features were detected either in EUV or soft X-ray. Here we show two examples of these sources in Fig. \ref{fig:network_sources}. The source in the left panel is detected both in the EOVSA and VLA, while the source in the right panel is detected in multiple VLA bands. Hence both of these sources are high-fidelity sources and hence chosen for further analysis using the EUV and X-ray data. In Fig. \ref{fig:network_sources} the details of these two sources are shown in the same format as that in Fig. \ref{fig:coronal_hole_sources}. In the top left panel, the VLA and EOVSA data are shown with blue and white contours respectively. In the top right panel, blue, white, and magenta contours correspond to images at 1.057, 1.313 and 1.441 GHz respectively. Dashed cyan circles have been drawn such that their centers show the location from where the EUV light curves shown in the bottom panel are extracted. {Movies showing the EUV variability at the location of these sources are provided in the supplementary material. The format of these movies is the same as that in Figs. \ref{fig:spw2_coronal_hole_source1}, \ref{fig:spw7_coronal_hole_source1} and \ref{fig:spw7_coronal_hole_source2}. Still frames of the movies are provided in Figs. \ref{fig:spw5_network1} and \ref{fig:spw2_network1}.} For the source in the left panel, we see a clear peak in the X-ray light curve at the same time as the peak in the radio. For the source in the right panel, while we have marked the numerical peak of the radio light curve, the fluxes are comparable at the times when the source was detected. Additionally, there are multiple X-ray peaks during this time interval, making it hard to identify a probable X-ray counterpart to the radio source. However, as mentioned earlier due to the high contamination from electrons, these signatures in the GOES lightcurve should not be treated as confirmatory signatures.

\section{Discussion}

In this paper, we take advantage of simultaneous broadband radio imaging spectroscopy observations made with the EOVSA and VLA to detect and analyze weak (a few times 0.01 sfu) radio transients from the quiescent solar corona in the microwave range. 
Reports of such microwave transients are very rare in the literature, and our study marks the first in which they are detected in multiple frequencies simultaneously. With these measurements, we are able to pinpoint the emission mechanisms and show that they are generally consistent with a nonthermal gyrosynchrotron origin. The nonthermal nature is also true for one radio transient event associated with a CBP (discussed in Section \ref{sec:cbp}). We note that, however, in the past works, the radio counterparts of CBPs were usually attributed to the free-free emission mechanism \citep[see, e.g.,][and references therein]{madjarska2019}. We caution that, due to the small number of sources studied here, one can not attribute most of the radio transients at these frequencies to nonthermal emission. Nevertheless, the detection of gyrosynchrotron emission from these small scales transients does provide strong evidence of the presence of nonthermal electrons in the quiescent solar corona. The nonthermal electron energies estimated in this work indicate that these nonthermal electrons are either comparable to or weaker than those detected even during regular flares and microflares \citep{glesener2020}. These results are also consistent with the recent detection of WINQSEs, which are hypothesized to be produced due to coherent plasma emission from very weak nonthermal electron beams in the quiet solar corona \citep{mondal2020,sharma2022,mondal2023} and X-ray detection of quiet sun transients \citep{glesener2020, cooper2021, paterson2022}.

In Section \ref{sec:coronal_hole_sources} we present robust detection of radio transients from CBPs lying in coronal holes. To the best of our knowledge, this is the first detection of such sources in microwave wavelengths. In one instance, we have also shown that the spectrum is inconsistent with a free-free emission origin and may be explained by a nonthermal gyrosynchrotron model. The presence of nonthermal emission implies the occurrence of weak energy release events in the coronal hole region due to yet-to-determined mechanisms such as interchange reconnection \citep[e.g.][]{fisk2001,upendran2022}. 

Measuring the magnetic field in the solar corona has always been very difficult and routine measurements of this nature are not possible yet. While there are techniques to measure the magnetic field using EUV, infrared, and optical data, these are limited to regions beyond the solar disc \citep{casini2017}. On the other hand, most radio measurements reported till date has only measured the magnetic field from active regions, or, are associated with explosive events like solar flares and coronal mass ejections \citep{alissandrakis2021}. To the best of our knowledge, this work is the second work that constrains the coronal magnetic field in the quiet on-disk solar corona using radio observations (the first one was done by \citet{habbal1986}). The availability of more sensitive and wideband radio instruments thus raises the hope that this technique of using quiet sun radio transients can be used in a much more routine manner to constrain the magnetic field in the quiescent solar corona. 

\section{Conclusion}

In this work, we have presented a study on weak radio transients from the quiet sun in the microwave range. We take advantage of imaging spectroscopy observations made by VLA and EOVSA along with available EUV data to unravel features of some of these transients. We show that at least 3 of the detected radio transients observed at 1--2 GHz are powered by nonthermal gyrosynchrotron emission. We have modeled the spectrum using a gyrosynchrotron model whenever possible and using the modeled parameters to estimate the nonthermal electron energy, which varies between $10^{26}-10^{29}\,$ergs. While this result is obtained from only a handful of events, our results strongly suggest the presence of a significant nonthermal electron population even during these quiet times.  For the two sources where we were able to calculate both thermal and nonthermal energy, we find that they are comparable. This suggests that the nonthermal energy budget should also be taken into account when calculating the total coronal energy budget.

We also present the first detection of radio transients associated with coronal holes and show that at least in one instance the source is due to gyrosynchrotron emission. We suggest that the source may be due to interchange reconnection in the vicinity of coronal holes, a scenario suggested to explain solar wind formation from the coronal holes. 

In addition, this work also shows that these radio transients can be used to constrain the coronal magnetic field in the quiet sun. Recent advances with the EOVSA have already demonstrated the technique of using microwave imaging spectroscopy to constrain the coronal magnetic field based on nonthermal gyrosynchrotron radiation theories \citep[e.g.][]{chen2020,kuroda2020,yu2020,wei2021,chen2021,fleishman2022,zhang2022}. However, due to the limit in sensitivity, dynamic range, and image fidelity of current instruments, such studies have largely been limited to active regions and solar flares. We hope that with the availability of more sensitive and broadband instruments with a much lower sidelobe level and higher image fidelity, similar studies can be extended even to the quiet solar corona. While due to the isolated nature of these radio transients, this technique can not be used to map the coronal magnetic field of the full solar disk, these measurements will add new constraints which we hope would ultimately lead to a better understanding of the magnetic field structure of the solar corona.

\begin{acknowledgments}
This work makes use of public VLA data from the observing program VLA/19B-338. The NRAO is a facility of the National Science Foundation (NSF) operated under a cooperative agreement by Associated Universities, Inc. The authors acknowledge Dr. Tim Bastian for leading the VLA observing program and helpful discussions. SM also acknowledges Dr. Stephen White for discussions regarding intricacies of the GOES data.  This work is supported by NSF grant AGS-1654382 
and NASA grants 80NSSC20K0026, 
80NSSC20K1283,  
and 80NSSC21K0623 
to NJIT. EOVSA operations are supported by NSF grants AST-1910354 and AGS-2130832 to NJIT.
This research used version 4.0.5 \citep{mumford_2022_zenodo} of the SunPy open source software package \citep{sunpy_community2020}.
This research used version 0.6.4 \citep{barnes2020b} of the aiapy open source software package \citep{Barnes2020}.
This research has made use of NASA's Astrophysics Data System Bibliographic Services.
\end{acknowledgments}

 \appendix

 \section{VLA Data analysis and Imaging procedure}

 All analysis were done using the Common Astronomy Software Applications \citep[CASA,][]{mcmullin2007}. 
 A combination of manual flagging and an automated flagging routine, \textit{tfcrop}, was used for the initial flagging. Calibration was non-trivial as the phase calibrator becomes resolved at baselines above 3$k\lambda$, a threshold below which many antennas had very few baselines. Hence to determine robust and accurate antenna gains self-calibration technique was applied on the phase calibrator data. We initially calibrated the data using baselines smaller than $\sim 5k\lambda$ at 1.05 GHz (translates to a scale of $\sim 40^{''}$) and used progressively higher thresholds at higher frequencies so as to ensure that sufficient baselines were present for all antennas within this threshold. Each spectral window was calibrated independently assuming that the calibrator source flux is constant within each spectral window. After the initial calibration of bandpass and temporal variability of gain using baselines below the chosen threshold, the obtained gains were refined using a self-calibration based approach. During the self-calibration approach, all baselines were used. We also assumed that the bandpass obtained earlier using the uv-based threshold is correct and time-invariant. Flagging was done using the task \textit{rflag} when needed. After the self-calibration converged, we fitted a Gaussian function to the source using the task \textit{imfit} and obtained a scaling factor to match the observed flux and true flux obtained from the VLA calibrator manual.  The obtained gains from the phase calibrator and complex gains due to the 20 dB attenuators were then applied on the solar data\footnote{See \citet{Chen2013} for a more detailed discussion on the use and calibration of the 20 dB attenuators.}. A round of flagging was done using the task \textit{rflag} to remove bad data in solar scans. 

Imaging was done using the task \textit{tclean}. We have used the auto-masking mode for deconvolving only ``real" sources. However, the point spread function, or synthesized beam, generally had a sidelobe level of about 40\%, which was larger than that typically observed in full-day synthesis images. Hence different auto-masking parameters of the task \texttt{tclean} were tuned so that the emission could be better captured. Specifically, \texttt{sidelobethreshold}, \texttt{noisethreshold}, \texttt{minbeamfrac}, and \texttt{cyclefactor} were set to 1.5, 3, 0.2, and 5, respectively\footnote{Detailed description of these parameters are provided in \url{https://casadocs.readthedocs.io/en/stable/notebooks/synthesis_imaging.html}}.
 
\software{Numpy \citep[][]{Harris2020},  
          Scipy \citep[][]{Scipy2020},
           Python 3 \citep[][]{python3},
           Matplotlib \citep[][]{Hunter:2007},
           Sunpy \citep[][]{sunpy_community2020},
           Astropy\citep{astropy2013,astropy2018,astropy2022},
           Aiapy\citep{barnes2020b}
           }

\facilities{OVRO:SA, SDO, GOES}
           
\bibliography{sample631}{}
\bibliographystyle{aasjournal}

\end{document}